# Interpreting the Suction Stress Profiles under Steady-State Conditions Considering the Independence of van Genuchten SWCC Parameters

Sumanta Roy[1] and Manash Chakraborty[2*]

[1] Dept. of Civil Engineering, Indian Institute of Technology Madras, Chennai-600036, India

[2] Civil Engineering Department, Indian Institute of Technology (BHU), Varanasi-221005, India

Emails: roysumanta47@gmail.com; manashchakra.civ@itbhu.ac.in

## Abstract

Suction stress is a fundamental component for applying the effective stress principle in unsaturated geotechnical engineering problems. The present paper aims to understand how the suction stress profiles get influenced if the *m* and *n* parameter of the van-Genuchten SWCC model is completely independent. Through the analysis, it is well noted that for constant air entry value, if the *m* and *n* relationships are varied, not only the number and the spread of the characteristic regimes get influenced, but the nature and the trend of the suction stress profiles associated with the regimes also alter significantly. A suggestion of selecting the *m* and *n* relationships is proposed for obtaining conservative solutions for geotechnical stability problems. The relevance of considering the unconstrained *m* parameter is established by computing the tensile crack depth of the clayey slope under steady-state infiltration conditions.

**Keywords** Suction stress; SWCC; HCF; SSCC; Steady state condition

* For Correspondence

## 1. Introduction

The concept of effective stress, as pioneered by Terzaghi [1, 2], was a major breakthrough in developing the field of soil mechanics. Effective stress is the single stress state parameter to govern the load-bearing capacity and the deformational aspects of two extreme states of the soil - completely saturated and completely dry. In the early sixties of the previous century, researchers [3-5] put their effort into developing and applying the effective stress principle for unsaturated soil medium. However, unlike saturated soil mechanics, there are two widely accepted independent stress state variables that describe the volume change, distortion, shear strength, and other physical behaviour of the unsaturated soil medium. Bishop's [3] effective stress equation is one of the oldest and frequently used equations that relates net normal stress $(\sigma - u_a)$ to matric suction $(u_a - u_w)$

$$\sigma' = (\sigma - u_a) + \chi (u_a - u_w) \tag{1}$$

Here, $\sigma'$ is the effective stress; $\chi$ = co-efficient of effective stress is a single-valued constitutive property, which is related to the degree of saturation and lies between 0 (dry soils) and 1 (fully saturated soils). But Bishop's effective stress equation suffers serious limitations due to lack of theoretical basis, low impact on practical implementation, and difficulty in experimental findings. Fredlund [6, 7] rightly pointed out that "Bishop's equation should not be referred to as a fundamental description of stress state" and be referred to as a constitutive equation. Experiments showed that the parameter $\chi$ can go beyond the range of 0-1 and is not consistent as it is different for volume change behaviour and shear strength prediction. Moreover, the term $\chi$ is enforced to zero when the soil is relatively in a dry state. This is because of the fact that the term $\chi\left(u_a - u_w\right)$ does not account for the physicochemical forces, (such as, van-Der Waals force, electrical attraction/ repulsion forces, double diffusion repulsive force, and chemical forces) which could be as high as several hundred kilopascals, in the presence of fines, even when the soil is dry. Later, in the early twenty-first century, the concept of suction stress was established by Prof. Lu and his co-workers [8-10]. The suction stress ($\sigma^s$) is referred to as a stress variable and it takes into account of all the physical interparticle stress mechanisms, such as physicochemical forces, surface tension, and capillarity. Considering the suction stress, Lu and Likos [8] proposed a new unified effective stress equation irrespective of soil type and saturation state

$$\sigma' = \sigma - u_a - \sigma^s \tag{2}$$

A closed-form expression of suction stress for non-negative matric suction was further recommended by Lu *et al.* [11] based on the thermodynamic equilibrium principle

$$\sigma^s = -\left(u_a - u_w\right)\Theta_n = -\left(u_a - u_w\right)S_e \tag{3}$$

It is to be noted that, in terms of magnitude, $\Theta_n$ (normalized water content) and $S_e$ (effective degree of saturation) are the same, as shown in Appendix- A.

A few empirical soil water characteristics curves (SWCC) are proposed for describing soil's constitutive relation between the normalized variables ($\Theta_n$, $S_e$) and the matric suction [12-16]. The following three-parameters-dependent SWCC model, as proposed by van Genuchten [13], is one of the most popular and widely adopted models for solving various problems in unsaturated soil mechanics:

$$\Theta_n = S_e = \frac{1}{\left\{1+\left[\alpha\left(u_a-u_w\right)\right]^n\right\}^m} \tag{4}$$

Here, $\alpha$, $n$ and $m$ are the fitting parameters of the SWCC curve; $\alpha$ , in a sense, reflect the inverse of air entry value (AEV), $n$ indicates the dimensionless pore spectrum number, and $m$ controls the sigmoidal shape of the SWCC curve. Precisely $\alpha^{-1}$ is somewhat greater than the actual AEV, $n$ provides the desaturation rate of the soil as $\psi > \psi_{AEV}$, and $m$ gives greater flexibility in fitting SWCC data. Combining equations (2) and (3), a unique closed-form equation of soil suction characteristics curve (SSCC) is derived for all type of soils [11]:

$$\sigma^s = -\left(u_a-u_w\right)\left\{\frac{1}{1+\left[\alpha\left(u_a-u_w\right)\right]^n}\right\}^m \tag{5}$$

The SSCC is intrinsically related with SWCC and owing to the existence of unique relation between SWCC and SSCC, same set of material parameters are used for defining both the curves.

However, following van Genuchten – Mualem [17] proposition, $m$ is assumed to bear the following fixed relationship with $n$: $m$=1-1/$n$. This relation is used in many theoretical developments [18-22] and practical applications [23-26]. Based on this special relationship between $m$ and $n$, Lu and Griffiths [18] developed a theoretical basis for predicting profiles of suction stress in unsaturated soils considering a wide range of hydromechanical parameters and evaporative/infiltrative conditions.

In this article, the interdependence of $m$ and $n$ are relaxed and an effort has been made to develop a generalized theory to predict suction stress profiles without providing any restrictions to the van Genuchten (vG) parameters – $m$ and $n$. Each correlation of $m$ and $n$ parameters would result in a different suction stress profile and it is extremely important to choose the proper values of the vG parameters for safe analysis of various engineering projects. Well-defined regimes are identified in the three-dimensional space of model and flow parameters. The suction stress profiles pertaining to each regime are definite and possess distinct features. Owing to its direct influence on effective stress, the proper knowledge of suction stress profiles is quite necessary for implementing in various analytical and numerical techniques.

## 2. Problem definition

A partially saturated zone, as shown in figure 1, is lying above the water table ($z$=0). On account of seasonal variations or other external influences (e.g. occasional floods, transpiration of plants, watering of lawns) water is likely to flow within the soil. Typical forms of matric suction are to be plotted for evaporative, infiltrative, and no-flow characteristics. It is intended to comprehensively study how the flow parameters and the unsaturated soil properties combinedly affect the suction stress characteristics. In this analysis, the hysteresis of the SWCC curves is not considered explicitly. Hence, the predictions are valid solely for monotonic wetting or monotonic drying but not for composite processes where wetting and drying occur simultaneously or sequentially. However, the obtained results are deliberately plotted in terms of nondimensional factors. Therefore, the AEV related to drying and wetting can be taken separately.

## 3. Suction stress formulation and associated assumptions

(a) The one-dimensional, vertical, steady state seepage flow in unsaturated soil medium is modelled by employing Darcy's linear flow law, as under:

$$q=-k\left(\frac{d\left(u_a-u_w\right)}{\gamma_w dz}+1\right) \tag{6}$$

Here, osmotic, electrical, thermal, and velocity heads are considered to be negligible and hence, the fundamental driving potential for the water flow is the hydraulic head gradient. In this expression, $q$, $k$, and $\gamma_w$ indicates specific discharge, matric suction dependent unsaturated hydraulic conductivity, and unit weight of water, respectively.

(b) Owing to its simplicity and wide usage, the following Gardner's [15] one-parameter hydraulic conductivity function (HCF) is employed to relate the characteristic dependency of hydraulic conductivity on matric suction

$$k=k_s e^{-\alpha\left(u_a-u_w\right)} \tag{7}$$

here, $k_s$ refers to saturated hydraulic conductivity. The physical consequence of this exponential equation is that the unsaturated hydraulic conductivity decays significantly with the increase in the matric suction and this reduction is quite rapid for sandy soils.

(c) By substituting Gardner's HCF into Darcian flow equation, equation (6) turns into the following form:

$$q=-k_s e^{-\alpha\left(u_a-u_w\right)}\left(\frac{d\left(u_a-u_w\right)}{dz}+1\right) \tag{8}$$

(d) Integrating the above equation and imposing the boundary condition of zero suction at the water table leads to the following expression:

$$u_a - u_w = -\frac{1}{\alpha}\ln\left[\left(1+\frac{q}{k_s}\right)e^{-\alpha\gamma_w z} - \frac{q}{k_s}\right] \tag{9}$$

(e) Plugging the expression of equation (9) in equation (5), the suction stress can be rewritten in the following dimensionless form:

$$\alpha\sigma^s = \frac{\ln\left[\left(1+\frac{q}{k_s}\right)e^{-\gamma_w\alpha z} - \frac{q}{k_s}\right]}{\left(1+\left\{-\ln\left[\left(1+\frac{q}{k_s}\right)e^{-\gamma_w\alpha z} - \frac{q}{k_s}\right]\right\}^n\right)^m} \tag{10}$$

Here, $\alpha\sigma^s$, $\gamma_w\alpha z$, and $q/k_s$ are the nondimensional terms associated with suction stress, depth, and flow ratio parameters. This equation is smooth, continuous, and differentiable.

3.1 *Variation of Matric Suction and Suction Stress*

The impact of soil types and the flow rates on the matric suction profiles are investigated on an 8m thick layer under constant $q$ conditions. The chosen soils properties along with the flow characteristics and the relative matric suction profiles are presented in figure 2. For the sandy soil, the infiltrative and evaporative matric suction profiles remain coincident with its hydrostatic counterpart up to an elevation relatively far from the water table. For an instance, the flow rates start to influence the matric suction profile of sand medium at a height of approximately 4.8 m above the water table. On the other hand, for silts and clays, the deviation of the steady-state suction profiles from the hydrostatic line is quite prominent adjacent to the water table. The variation of the matric suction measured at the ground surface ($z$=8) between the evaporative and precipitative profiles varies by about, 3 kPa, 50 kPa, and 108 kPa for sand, silt, and clay layers, respectively. This shows that the matric suction profiles for clayey soils are markedly influenced by the flow conditions.

Further, combining equations (5) and (6) following expression of the suction stress is derived:

$$\sigma^s = -\frac{S_e}{\alpha}\left[S_e^{\left(\frac{1}{m}\right)} - 1\right]^{\frac{1}{n}} \qquad 0 \le S_e \le 1 \tag{11}$$

The variation of $\sigma^s$ with $S_e$ for different soil types are pictorially represented in figure 3. For a certain value of $\alpha$ and *n*, higher the *m* lower is the suction stress. The impact of *m* on the suction stress curves appears to be significantly high for clays followed by silts and then sands. The figure gives an indication that for specific $\alpha$ and *n*, there is a critical $S_e$ ($S_{e,critical}$) at which the suction stress becomes maximum. This $S_{e,critical}$ is significantly lower for fine grained soils and it further decreases with the increase in *m*.

3.2 *Consideration of m parameter from Mualem's and Burdine's HCF model*

Among a number of empirical and macroscopic models, the following hydraulic conductivity models proposed by Burdine [27] and Mualem [17] are widely used especially for relating the vG *m* and *n* parameters:

$$\text{Burdine's model:} \quad k_w = k_s S_e^2 \frac{\int_0^{S_e} \frac{dS_e}{\psi^2(S_e)}}{\int_0^1 \frac{dS_e}{\psi^2(S_e)}} = k_s S_e^2 \frac{\int_0^{S_e} \left(\frac{S_e^{1/m}}{1 - S_e^{1/m}}\right)^{2/n} dS_e}{\int_0^1 \left(\frac{S_e^{1/m}}{1 - S_e^{1/m}}\right)^{2/n} dS_e} = k_s S_e^2 \frac{g(S_e)}{g(1)} \tag{12a}$$

$$\text{Mualem's model: } k_w = k_s S_e^{0.5}\left[\frac{\int_0^{S_e}\frac{dS_e}{\psi(S_e)}}{\int_0^1\frac{dS_e}{\psi(S_e)}}\right]^2 = k_s S_e^{0.5}\left[\frac{\int_0^{S_e}\left(\frac{S_e^{1/m}}{1-S_e^{1/m}}\right)^{1/n} dS_e}{\int_0^1\left(\frac{S_e^{1/m}}{1-S_e^{1/m}}\right)^{1/n} dS_e}\right]^2 = k_s S_e^{0.5}\left[\frac{f(S_e)}{f(1)}\right]^2 \quad (12b)$$

These models are based on the assumptions that the HCF and the SWCC model have one-to-one correspondence with the pore-size distribution of the soil. It was possible to use the vG SWCC model in equations (12a) and (12b) because equation (4) is invertible and the matric suction can be expressed as the following function of effective saturation:

$$\psi(S_e) = \left(\frac{S_e^{1/m}}{1-S_e^{1/m}}\right)^{-1} \quad (13)$$

The closed form expression of the permeability function which is consistent with the integral form of Burdine's and Mualem's model can be obtained by reducing the three fitting parameters into two parameters through the following definite relationships.

vG-Burdine's proposition: $m = 1 - \frac{2}{n}$ $(0<m<1;\ n>2)$ (14a)

vG -Mualem's proposition: $m = 1 - \frac{1}{n}$ $(0<m<1;\ n>1)$ (14b)

These mathematical constraints were to ensure the attainment of the closed-form solutions. However, by back calculating the *m* and *n* parameters from the large number of test results retrieved from the SoilVision database it was observed that these expressions do not represent most of the experimental data [28]. Van Genuchten and Nielsen [29] performed an extensive experimental study and concluded that "restricting *m* and *n* parameters limits the flexibility of vG-SWCC model in describing retention data of several soils". This motivates revisiting the suction stress profiles without imposing such strict restrictions on *m* parameter. In the present article, a wide range of combinations of vG parameters are chosen, as shown in figure 4. The regimes and the corresponding $\sigma^s$ evolved from various combinations of vG SWCC and flow parameters are summarized in the subsequent sections.

## 4. Different regimes, its associated suction stress profiles, and interpretation thereof

In terms of shape, maxima, and asymptotes, there are certain distinct characteristics of the suction stress versus depth curves, which eventually give rise to different solution regimes. The specific characteristics of the suction stress profiles depend on the gradient of the suction stress function. The gradient form of the dimensionless matric suction in terms of *m* and *n* parameters is derived in Appendix B. The range of the flow parameter is elaborated in Appendix C.

4.1 *Regimes and its associated suction stress profiles from Mualem's proposition*

Corresponding to Mualem's HCF model, the nondimensional matric suction can be expressed as (Appendix B):

$$\left\{-\ln\left[\left(1+\frac{q}{k_s}\right)e^{-\gamma_w \alpha z}-\frac{q}{k_s}\right]\right\}^n=\frac{1}{(K+1)n-2};\ K=m-1+\frac{1}{n} \quad (15a)$$

$$\Rightarrow -\gamma_w \alpha z=\ln\left\{\frac{e^{-((K+1)n-2)^{-1/n}}+q/k_s}{1+(q/k_s)}\right\} \quad (15b)$$

For the existence of real solutions:

(i) The exponential term within the logarithm must not contain a complex term, leading to $(K+1)n > 2$.

(ii) The value inside the logarithm must be positive. Since, $\frac{q}{k_s} > -1 \;\; \therefore \frac{q}{k_s} > -e^{-((K+1)n-2)^{-1/n}}$

These two considerations generate different regimes in the two-dimensional space of flow ratio and pore spectrum. The suction stress profiles corresponding to each regime have unique features. Table 1 illustrate the extent of each regime for various *m* and *n* relationship. Based on the *K* value, different number of regimes seem to evolve. This is due to the fact that by virtue of the first constraint for the existence of a real solution, there exists a critical value of *K* ($K_{crit}$), which demarcates between the total number of evolved regimes. The $K_{crit}$, related to Mualem's HCF model, is determined as follows:

$$1.1(K_{crit}+1) = 2 \Rightarrow K_{crit} = 0.8182 \tag{16}$$

Here, the coefficient 1.1 arises in view of considering the minimum value of *n*, as mentioned in a few literature [13, 19].

If *K* lies between 0 and $K_{crit}$, four regimes will be generated; conversely, if $K > K_{crit}$, the number of regimes will reduce to 3. Figures 5 and 6 present the schematic diagram of different regimes with the variation of *K* parameter; figure 5 depicts the expanse of the regimes for the realistic value of *n,* and figure 6 demonstrates the asymptotic approach of the *A*-line by considering high values of *n*. In the $q/k_s - n$ space, the *A*-line originates from (0,*n*), (0,0) and ($q/k_s$,0) points if the value of *K* is less than, equal to, and greater than $K_{crit}$, respectively. As the value of *K* goes below $K_{crit}$, *n* increases, and conversely, as *K* moves beyond $K_{crit}$, $q/k_s$ decreases.

By varying the relationship between *m* and *n*, figures 7-10 present the suction stress profiles for different regimes corresponding to two different scenarios, namely, (a) varying flow conditions with constant $n$ and (ii) varying $n$ with constant flow ratio; figures 7 and 8 corresponds to $K$ (=0 and 0.5) < $K_{crit}$, whereas, figures 9 and 10 pertains to $K$ (=1 and 2) > $K_{crit}$. Suction stress profiles illustrate the variation of the nondimensional suction stress with the nondimensional depth.

4.1a Regimes 1 and 2

It is well observed from the figures that there exists a prominent peak point on the suction stress profiles generated from Regimes 1 and 2. This is because of the existence of the real solutions of the equation $f'(z)=0$ in these two Regimes, as shown in Appendix B. The expressions of the maximum suction stress and its vertical position (measured from the water table) are mentioned in table 1 for various $K$ parameters. The maximum value of suction stress depends on $n$ and $K$, but not on the magnitude and direction of the water flow. However, the location of the peak suction stress is mainly related to $K$ and $q/k_s$ ratio. With the increase in $n$, the occurrence of the peak suction stress becomes quite prominent, especially in Regime 1. The bandwidth measured between the upward and downward limb of the suction stress profile significantly narrows down with the increase in $n$ and $q/k_s$ ratio. Moreover, as $K$ increases, the maximum suction stress decreases, and the occurrence of peak happens to be closer to the water table.

The qualitative difference of the suction stress profile in Regimes 1 and 2 is observed after the maximum value is surpassed. After the attainment of peak suction stress, the profiles corresponding to Regime 1 tend to zero within a certain depth above the water table, whereas, in the case of Regime 2, barring a few, most of the suction stress profiles asymptote towards a non-zero value as it approaches towards the ground surface. The asymptotic value of the suction stress profile is reported in the last row of table 1. The variation of this asymptotic value with $n$ is more significant when the pore size distributions are non-uniform, i.e., the impact in the asymptote appears remarkable when the change in $n$ is from 2 to 3 rather than 4 to 5. Moreover, with an increase in the $K$ parameter, the asymptotic value reduces; for instance, when $n$ =4 and $q/k_s$ =–0.3, the asymptotic value turns into 0.29 and 0.05, corresponding to $K$ = 0.5 and $K$ = 2, respectively.

4.1b Regimes 3 and 4

Unlike Regimes 1 and 2, the suction stress profiles generated from Regimes 3 and 4 do not possess pronounced peak points due to the emergence of the complex solution to the zero gradient of the function (i.e. $f'(z) = 0$). The suction stress profiles associated with Regime 3 monotonically increase up to a certain point and then asymptotes towards a constant value based on the $K$ parameter, $n$ value, and $q/k_s$ ratio. The expression of the asymptotic value remains to be the same as that given for Regimes 2. For a certain $q/k_s$, suction stress profiles from Regime 3 appear to be insensitive towards $n$ parameter up to a significant height above the water table; thereafter, the curves become completely distinct if $K < K_{crit}$. However, as the value of $K$ goes beyond $K_{crit}$, the pore size distribution in the soil matrix does not seem to impact the suction stress profiles associated with Regime 3. For a high percolation rate, if $K$ is greater than 1, unique suction stress profile evolves, irrespective of $n$. Contrary to Regime 2, an increase in steady-state infiltrative rate enforces the magnitude of the suction stress profiles of Regime 3 towards zero.

Regime 4 is indicative of a small infiltration situation. The suction stress profiles from Regime 4 approaches towards infinity as $\gamma_w \alpha z \to \ln(1 + k_s / q)$. It can be inferred that in case of Regime 4, the soil portion up to which the suction stress can be defined in the analysis solely depends on evaporative rate; $n$ and $K$ seem to be completely non-influential. For a higher evaporative rate, the suction stress turns into infinity within a small height above the water table. The high dependency of the suction stress profile on the evaporation rate is also pointed out earlier by a few researchers [30-32]. It is well noticed from figure 8(h) that the variation of the suction stress profile is quite notable when $q/k_s$ varies from 0.20 to 0.40, but when the flow rate varies from 0.80 to 0.95, there is an indiscernible change in the suction stress profile.

4.2 *Regimes and its associated suction stress profiles related to Burdine's HCF model*

Following Burdine's HCF model, the matric suction can be expressed as the following form (Appendix B):

$$\left\{-\ln\left[\left(1+\frac{q}{k_s}\right)e^{-\gamma_w \alpha z}-\frac{q}{k_s}\right]\right\}^n=\frac{1}{(K+1)n-3};\quad K=m-1+\frac{2}{n} \tag{17a}$$

$$\Rightarrow -\gamma_w \alpha z=\ln\left\{\frac{e^{-((K+1)n-3)^{-1/n}}+q/k_s}{1+(q/k_s)}\right\} \tag{17b}$$

Similar to Section 4.1, real solutions will exist, if: (i) $(K+1)n>3$ and (ii) $\dfrac{q}{k_s}>-e^{-((K+1)n-3)^{-1/n}}$.

Considering the minimum value of $n$, $K_{crit}$ is evaluated as follows:

$$1.1(K_{crit}+1)=3 \qquad \Rightarrow K_{crit}=1.7272 \tag{18}$$

Similar to Mualem's model, Burdine's HCF model generates either four ($K < K_{crit}$) or three ($K \geq K_{crit}$) different Regimes, as shown in figure 11. With an increase in the $K$ parameter, one can observe the following two occurrences: (a) diminishing of Regime 4 and simultaneous enhancement of Regime 1 and (b) shifting of origin of $A$-line from $(0, n) \rightarrow (0, 0) \rightarrow (q/k_s, 0)$.

Figures 12 and 13 display the suction stress profile corresponding to two different *K* parameters, namely, $K$=0 and $K$=1. Unlike Regimes 3 and 4, the suction stress profiles developed for Regimes 1 and 2 exhibit the peak characteristics. Corresponding to different *m* and *n* relations, table 2 displays the range of the Regimes, the expression of the maximum value of the suction profiles, and their location above the water table. For a specific *K*, the suction stress profiles from Regime 1, considering Burdine's proposition, advanced towards zero at a greater height than the curves evolved from Mualem's *m* and *n* relations. The suction stress profiles pertain to Regime 3 asymptote to the ground surface with finite suction stress, whereas, on the contrary, based on flow ratio, the suction curves related to Regime 4 become undefined after a certain portion above the water table. For higher *K*, suction stress profiles related to Regime 3 depend neither on *n* nor on *K* parameters.

4.3 *Three-Dimensional Representation*

Figure 14 represents the three-dimensional plot of the Regimes in $q/k_s-m-n$ space. The range of $q/k_s$ is selected between −1 and 1, whereas the range of *m* and *n* are selected between 0 to 10 and 1.1 to 9, respectively. The 3-D plot is shown from two different viewpoints to obtain a better perspective of the curved surface representing the boundaries between Regimes 1 and 4 (C surface) and Regimes 2 and 3 (B surface). The boundary between Regimes 1 and 2 (A surface) is a horizontal plane passing through $q/k_s = 0$ irrespective of *m* and *n* values. In a way, this plane demarcates between the infiltrative and evaporative zones. Further, the projection of the C surface on the floor of $m-n$ space results in 'J' shaped curves; with the increase in flow ratio, these curves continuously shift away from the $m$=0 axis. A narrow trough is observed on the B surface. As the value of $n$ increases, the projection of the B-surface on the $q/k_s$–$m$ plane falls with the increase in $n$ value.

4.4 *Summary of the importance of m and n relationship*

The variation of the area ratio with various $m$ and $n$ relations are presented in figure 15; here, the area of a certain Regime corresponding to a specific $K$ is normalized with respect to the area of that Regime pertaining to $K$=0. It is to be noted that these areas are calculated while $n$ lies between 1.1 and 8. The curves are plotted for the four Regimes evolved from various $K$ parameters of Mualem's and Burdine's HCF models. It can be well noted that with the increase in the $K$ parameter, the area encompassed by Regimes 1 and 2 grows in size, whereas the area enclosed by Regimes 3 and 4 shrinks. Nevertheless, all the curves approach the horizontal plateau, indicating that after a certain $K$ (almost $K$=4.5), any further change in $K$ parameter ceases to impact the trend of the curve evolved from any Regime. The rate of area decrement for Regime 4 is higher in case of Burdine's HCF models. The area ratio curve generated from Regimes 3 almost coincide for both the HCF models; however, the area ratio curves pertain to Regimes 1 and 2 are on the higher side if it is computed by Burdine's HCF model.

Figure 16 indicate the suction stress profiles above the water table for various $m$ and $n$ relations evolved from Burdine's and Mualem's HCF models. The profiles appear to be independent of $K$ parameters up to a certain height till the maximum suction stress is reached; beyond that, the relationship of $m$ and $n$ significantly influence the suction stress profiles. With the increase in $K$, (a) the deviations between the suction stress profiles decreases, especially for Burdine's $K$ parameter and (b) the peak point in the suction stress profiles becomes smaller and closer to the water table. The influence of $K$ is significantly high for the curves originating from Regime 2. However, the $K$ parameter not at all influence the suction stress profiles generated from Regime 3. This discussion shows that choosing high values of $K$ (maybe 3 or 4) will results in lower prediction of the geotechnical stability problems, e.g., bearing capacity of foundations, different earth pressure problems or the safety factors for soil slopes.

Figures 17(a) and 17(b) exclusively depicts how the absolute magnitude of the maximum suction stress varies for various combinations of $n$ and $K$ parameters. On these curves, peak points are viewed for lower values of $K$ (e.g. 0, 0.5 for Mualem's and 0, 1.0 for Burdine's HCF model); whereas, for higher $K$'s these curves continuously increase. Along with the magnitude, the position at which the maximum suction stress is recorded explicitly varies with $K$ parameter. Thus, it can be summarized that the relationship between vG $m$ and $n$ parameters extensively impact (a) the number and extent of the Regimes, (b) the magnitude and the occurrence of the maximum suction stress, and (c) the suction stress profiles especially for Regimes 1 and 2.

## 5. Verification

For establishing the necessity of considering independent $m$ parameter, the tensile crack depth of clayey slope is numerically computed from the following implicit expression as provided in the work of Sun *et al.* [33].

$$z_c = \frac{1-2\mu}{\gamma\mu\alpha} \frac{-\ln\left[\left(1+\frac{q}{k_s}\right)e^{-\gamma_w\alpha(h_w-z_c)}-\frac{q}{k_s}\right]}{\left(1+\left\{-\ln\left[\left(1+\frac{q}{k_s}\right)e^{-\gamma_w\alpha(h_w-z_c)}-\frac{q}{k_s}\right]\right\}^n\right)^m} \tag{19}$$

Here, $h_w$ and $z_c$ are the water table depth and the crack depth measured from the ground surface, respectively. The solutions are presented in terms of normalized crack depth. Figure 18 shows the variation of with respect to $h_w$ corresponding to various *m* and *n* relationships. There is an inset in the figure to show the schematic representation of the analyzed problem. The curves provide a clear impression that no matter whatever way *m* and *n* are related, as long as there is no flow in the unsaturated medium, the depth of tension crack remains unchanged. However, with the increase in steady state infiltration rate the crack openings gradually closes up and the amount of closure depends on *m* values. Higher values of *K* estimate lower crack depth. Moreover, Burdine's HCF model predict larger crack openings than its Mualem's counterpart. This study reveals the importance of treating the *m* parameter without imposing any constraints.

## 6. Conclusions

This article primarily focuses on how the form of the unified effective stress above the water table gets impacted if two of the curve fitting parameters of van Genuchten's SWCC model, namely, *m* and *n* are entirely independent. In this regard, a generalized theory is proposed to predict the suction stress profiles without providing any kind of restrictions on the exponent parameter. For this purpose, a number of different *m* and *n* relationships are employed by following the hydraulic conductivity models proposed by Mualem and Burdine. It is well observed that imposing a strict restriction on the relationship of *m* and *n* parameters, as provided in the literature, may lead to excessive overprediction of the stability values. Corresponding to the three flow conditions, namely, no-flow, infiltrative, and evaporative, a suggestion is made for the practicing engineers to choose the *m* and *n* relation in order to obtain conservative solutions to classical geotechnical stability problems.

## Appendix A

Normalized volumetric water content $\Theta$ is given by: $\Theta_n = \dfrac{\theta - \theta_r}{\theta_s - \theta_r}$

$$\Theta_n = \frac{\dfrac{V_w}{V_s} - \dfrac{V_{w,residual}}{V_s}}{\dfrac{V_{w,saturated}}{V_s} - \dfrac{V_{w,residual}}{V_s}} = \frac{V_w - V_{w,residual}}{V_{w,saturated} - V_{w,residual}} = \frac{\dfrac{V_w}{V_V} - \dfrac{V_{w,residual}}{V_V}}{\dfrac{V_{w,saturated}}{V_V} - \dfrac{V_{w,residual}}{V_V}} = \frac{S - S_r}{1 - S_r} = S_e \qquad \text{(A.1)}$$

($\theta, \theta_s$, and $\theta_r$), ($V_w, V_{w,saturated}$, and $V_{w,residual}$), and ($S$, 1, and $S_r$) are the set of the volumetric water content , volume of water, and degree of saturation at any arbitrary, saturated, and residual state, respectively.

## Appendix B

For a sustained flow ratio and constant SWCC model parameters equation (10) provides the expression of nondimensional suction stress as function of spatial coordinate:

$$\alpha\sigma^s = \frac{\ln\left[\left(1 + \dfrac{q}{k_s}\right)e^{-\gamma_w \alpha z} - \dfrac{q}{k_s}\right]}{\left(1 + \left\{-\ln\left[\left(1 + \dfrac{q}{k_s}\right)e^{-\gamma_w \alpha z} - \dfrac{q}{k_s}\right]\right\}^n\right)^m} = f(z) \qquad \text{(B.1)}$$

For finding the extremum on the suction stress profiles,

$$f'(z) = [\alpha\sigma^s]' = 0 \qquad \text{(B.2)}$$

$$f'(z) = \frac{[\alpha h]'\left\{1 + (\alpha h)^n\right\}^m - \ln\left[\left(1 + \dfrac{q}{k_s}\right)e^{-\gamma_w \alpha z} - \dfrac{q}{k_s}\right] m\left\{1 + (\alpha h)^n\right\}^{m-1} (\alpha n)(\alpha h)^{n-1} h'}{\left(1 + (\alpha h)^n\right)^{2m}} = 0 \qquad \text{(B.3)}$$

Here, $h$ = Matric Suction ($h = u_a - u_w$). Equation (B.3) leads to

$$\alpha h'\left\{1+(\alpha h)^n\right\}^m = \ln\left[\left(1+\frac{q}{k_s}\right)e^{-\gamma_w \alpha z} - \frac{q}{k_s}\right] m\left\{1+(\alpha h)^n\right\}^{m-1}(\alpha h)(\alpha h)^{n-1} h'$$

$$\Rightarrow\ 1+(\alpha h)^n = \ln\left[\left(1+\frac{q}{k_s}\right)e^{-\gamma_w \alpha z} - \frac{q}{k_s}\right](mn)[\alpha h]^{n-1}$$

$$\Rightarrow\ \frac{1+(\alpha h)^n}{mn[\alpha h]^{n-1}} = \ln\left[\left(1+\frac{q}{k_s}\right)e^{-\gamma_w \alpha z} - \frac{q}{k_s}\right]$$

$$\Rightarrow\ \frac{1+(\alpha h)^n}{mn\left\{-\ln\left[\left(1+\frac{q}{k_s}\right)e^{-\gamma_w \alpha z} - \frac{q}{k_s}\right]\right\}^{n-1}} = \ln\left[\left(1+\frac{q}{k_s}\right)e^{-\gamma_w \alpha z} - \frac{q}{k_s}\right]$$

$$\Rightarrow\ 1+(\alpha h)^n = mn\left\{-\ln\left[\left(1+\frac{q}{k_s}\right)e^{-\gamma_w \alpha z} - \frac{q}{k_s}\right]\right\}^n$$

$$\Rightarrow\ 1+\left\{-\ln\left[\left(1+\frac{q}{k_s}\right)e^{-\gamma_w \alpha z} - \frac{q}{k_s}\right]\right\}^n = mn\left\{-\ln\left[\left(1+\frac{q}{k_s}\right)e^{-\gamma_w \alpha z} - \frac{q}{k_s}\right]\right\}^n$$

$$\Rightarrow\ \frac{1}{mn-1} = \left\{-\ln\left[\left(1+\frac{q}{k_s}\right)e^{-\gamma_w \alpha z} - \frac{q}{k_s}\right]\right\}^n \tag{B.4}$$

Considering Mualem's and Burdine's $K$ parameter, equation (B.4) turns into the following form

$$\Rightarrow\ \frac{1}{mn-1} = \frac{1}{\left[(K+1)-\frac{1}{n}\right]n-1} = \frac{1}{(K+1)n-2} = \left\{-\ln\left[\left(1+\frac{q}{k_s}\right)e^{-\gamma_w \alpha z} - \frac{q}{k_s}\right]\right\}^n \tag{B.5}$$

$$\Rightarrow\ \frac{1}{mn-1} = \frac{1}{\left[(K+1)-\frac{2}{n}\right]n-1} = \frac{1}{(K+1)n-3} = \left\{-\ln\left[\left(1+\frac{q}{k_s}\right)e^{-\gamma_w \alpha z} - \frac{q}{k_s}\right]\right\}^n \tag{B.6}$$

## Appendix C

In nondimensional suction stress equation provided in equation (10), following two constraints are to be imposed on the bracketed quantity within the logarithm: (a) the quantity should be greater than zero according to mathematical perspective and (b) to ensure non-negative matric suction the quantity should be less than or equal to unity.

$$0 < \left(1 + \frac{q}{k_s}\right) e^{-\gamma_w \alpha z} - \frac{q}{k_s} \leq 1.0 \tag{C.1}$$

These two upper and lower constraints in eq. (C.1) can be further rewritten as the following simplified form:

$$\frac{-k_s e^{-\gamma_w \alpha z}}{e^{-\gamma_w \alpha z} - 1} < q \leq k_s \tag{C.2}$$

The lower bound restriction on $q$ is the necessary condition for the validity of equation (10); violation of this condition will enforce matric suction to be zero in order to ensure real solutions. The criterion of maximum $q$ dictates that the magnitude of flux (infiltrative or evaporative) should always be less than or equal to the saturated hydraulic conductivity.

## List of figures:

**Figure 12.** The variation of suction stress profiles in four characteristic regimes considering Burdine's $K$ parameter with $K$=0 and corresponding to (a-d) constant $q/k_s$ and variable $n$ and (e-h) constant $n$ and variable $q/k_s$.

**Figure 13.** The variation of suction stress profiles in four characteristic regimes considering Burdine's $K$ parameter with $K$=1 and corresponding to (a-d) constant $q/k_s$ and variable $n$ and (e-h) constant $n$ and variable $q/k_s$.

**Figure 14.** Characteristic regimes of suction stress profiles in a 3D domain space considering complete independence of vG $m$ and $n$ parameters shown from (a) View 1 and (b) View 2.

**Figure 15.** The variation of area ratio with respect to the change in $K$ parameter.

**Figure 16.** The variation of suction stress profiles for three different regimes with different $m$ and $n$ relationship corresponding to (a, c, and e) Mualem's and (b, d, and f) Burdine's $K$ parameters.

**Figure 17.** The variation of the (a-b) magnitude and (c-d) location of the peak point of the non-dimensional suction stress profile with respect to the $n$ value corresponding to various $K$ parameter related with (a, c) Mualem's and (b, d) Burdine's HCF model.

**Figure 18.** The variation of normalized crack depth in clayey slopes with respect to $h_W$ conforming to steady state infiltration conditions.

**List of tables:**

**Table 1.** Extent of various Regimes and the specific features of the suction stress profiles for various Mualem's $K$ parameter

| | | $K$=0: $n(1-m)=1$ | $K$=0.5: $n(2m-3)=-2$ | $K$=1: $n(2-m)=1$ | $K$=2: $n(2-m)=1$ | $K$=3: $n(3-m)=1$ | $K$=4: $n(4-m)=1$ |
|---|---|---|---|---|---|---|---|
| Extent of Regimes in 2-D | 1 | $\frac{q}{k_s}\geq 0$ and $n>2$ | $\frac{q}{k_s}\geq 0$ and $n>1.33$ | $\frac{q}{k_s}\geq 0$ and $n>1.10$ | $\frac{q}{k_s}\geq 0$ and $n>1.10$ | $\frac{q}{k_s}\geq 0$ and $n>1.10$ | $\frac{q}{k_s}\geq 0$ and $n>1.10$ |
| | 2 | $0>\frac{q}{k_s}>-e^{-(n-2)^{-1/n}}$<br>$n$>2 | $0>\frac{q}{k_s}>-e^{-(1.5n-2)^{-1/n}}$<br>$n$>1.33 | $0>\frac{q}{k_s}>-e^{-(2n-2)^{-1/n}}$<br>$n$>1.10 | $0>\frac{q}{k_s}>-e^{-(3n-2)^{-1/n}}$<br>$n$>1.10 | $0>\frac{q}{k_s}>-e^{-(4n-2)^{-1/n}}$<br>$n$>1.10 | $0>\frac{q}{k_s}>-e^{-(5n-2)^{-1/n}}$<br>$n$>1.10 |
| | 3 | (i) $-1<\frac{q}{k_s}\leq -e^{-(n-2)^{-1/n}}$<br>$n$>2 and<br>(ii) $\frac{q}{k_s}<0$; $1.1<n<2$ | (i) $-1<\frac{q}{k_s}\leq -e^{-(1.5n-2)^{-1/n}}$<br>$n$>1.33 and<br>(ii) $\frac{q}{k_s}<0$; $1.1<n<1.33$ | $-1<\frac{q}{k_s}\leq -e^{-(2n-2)^{-1/n}}$<br>$n$>1.10 | $-1<\frac{q}{k_s}\leq -e^{-(3n-2)^{-1/n}}$<br>$n$>1.10 | $-1<\frac{q}{k_s}\leq -e^{-(4n-2)^{-1/n}}$<br>$n$>1.10 | $-1<\frac{q}{k_s}\leq -e^{-(5n-2)^{-1/n}}$<br>$n$>1.10 |
| | 4 | $\frac{q}{k_s}\geq 0$ and $n<2$ | $\frac{q}{k_s}\geq 0$ and $n<1.33$ | Not exist | Not exist | Not exist | Not exist |
| $(\alpha\sigma^s)_{\max}$ Reg. 1&2 | | $\frac{(n-2)^{(n-2)/n}}{(n-1)^{(n-1)/n}}$ | $\frac{(1.5n-2)^{(1.5n-2)/n}}{(1.5n-1)^{(1.5n-1)/n}}$ | $\frac{(2n-2)^{(2n-2)/n}}{(2n-1)^{(2n-1)/n}}$ | $\frac{(3n-2)^{(3n-2)/n}}{(3n-1)^{(3n-1)/n}}$ | $\frac{(4n-2)^{(4n-2)/n}}{(4n-1)^{(4n-1)/n}}$ | $\frac{(5n-2)^{(5n-2)/n}}{(5n-1)^{(5n-1)/n}}$ |
| $(\alpha\gamma_w z)_{\max}$ Reg. 1&2 | | $\ln\left\{\frac{(1+q/k_s)e^{(n-2)^{-1/n}}}{1+\frac{q}{k_s}e^{(n-2)^{-1/n}}}\right\}$ | $\ln\left\{\frac{(1+q/k_s)e^{(1.5n-2)^{-1/n}}}{1+\frac{q}{k_s}e^{(1.5n-2)^{-1/n}}}\right\}$ | $\ln\left\{\frac{(1+q/k_s)e^{(2n-2)^{-1/n}}}{1+\frac{q}{k_s}e^{(2n-2)^{-1/n}}}\right\}$ | $\ln\left\{\frac{(1+q/k_s)e^{(3n-2)^{-1/n}}}{1+\frac{q}{k_s}e^{(3n-2)^{-1/n}}}\right\}$ | $\ln\left\{\frac{(1+q/k_s)e^{(4n-2)^{-1/n}}}{1+\frac{q}{k_s}e^{(4n-2)^{-1/n}}}\right\}$ | $\ln\left\{\frac{(1+q/k_s)e^{(5n-2)^{-1/n}}}{1+\frac{q}{k_s}e^{(5n-2)^{-1/n}}}\right\}$ |
| $(\alpha\sigma^s)_{z\to\infty}$ Reg. 2&3 | | $\frac{-\ln(-q/k_s)}{\left\{1+\left[-\ln\left(\frac{-q}{k_s}\right)\right]^n\right\}^{1-1/n}}$ | $\frac{-\ln(-q/k_s)}{\left\{1+\left[-\ln\left(\frac{-q}{k_s}\right)\right]^n\right\}^{1.5-1/n}}$ | $\frac{-\ln(-q/k_s)}{\left\{1+\left[-\ln\left(\frac{-q}{k_s}\right)\right]^n\right\}^{2-1/n}}$ | $\frac{-\ln(-q/k_s)}{\left\{1+\left[-\ln\left(\frac{-q}{k_s}\right)\right]^n\right\}^{3-1/n}}$ | $\frac{-\ln(-q/k_s)}{\left\{1+\left[-\ln\left(\frac{-q}{k_s}\right)\right]^n\right\}^{4-1/n}}$ | $\frac{-\ln(-q/k_s)}{\left\{1+\left[-\ln\left(\frac{-q}{k_s}\right)\right]^n\right\}^{5-1/n}}$ |

**Table 2.** Extent of various Regimes and the specific features of the suction stress profiles for various Burdine's $K$ parameter

| | | $K$=0: $n(1-m)=2$ | $K$=0.5: $n(3-2m)=4$ | $K$=1: $n(2-m)=2$ | $K$=2: $n(2-m)=2$ | $K$=3: $n(3-m)=2$ | $K$=4: $n(4-m)=2$ |
|---|---|---|---|---|---|---|---|
| Extent of Regimes in 2-D | 1 | $\frac{q}{k_s}\ge 0$ and $n>3$ | $\frac{q}{k_s}\ge 0$ and $n>2$ | $\frac{q}{k_s}\ge 0$ and $n>1.5$ | $\frac{q}{k_s}\ge 0$ and $n>1.10$ | $\frac{q}{k_s}\ge 0$ and $n>1.10$ | $\frac{q}{k_s}\ge 0$ and $n>1.10$ |
| Extent of Regimes in 2-D | 2 | $0>\frac{q}{k_s}>-e^{-(n-3)^{-1/n}}$ <br> $n>3$ | $0>\frac{q}{k_s}>-e^{-(1.5n-3)^{-1/n}}$ <br> $n>2$ | $0>\frac{q}{k_s}>-e^{-(2n-3)^{-1/n}}$ <br> $n>1.5$ | $0>\frac{q}{k_s}>-e^{-(3n-3)^{-1/n}}$ <br> $n>1.10$ | $0>\frac{q}{k_s}>-e^{-(4n-3)^{-1/n}}$ <br> $n>1.10$ | $0>\frac{q}{k_s}>-e^{-(5n-3)^{-1/n}}$ <br> $n>1.10$ |
| Extent of Regimes in 2-D | 3 | (i) $-1<\frac{q}{k_s}\le -e^{-(n-3)^{-1/n}}$ <br> $n>3$ and <br> (ii) $\frac{q}{k_s}<0$; $1.1<n<2$ | (i) $-1<\frac{q}{k_s}\le -e^{-(1.5n-3)^{-1/n}}$ <br> $n>2$ and <br> (ii) $\frac{q}{k_s}<0$; $1.1<n<2$ | (i) $-1<\frac{q}{k_s}\le -e^{-(2n-3)^{-1/n}}$ <br> $n>1.5$ <br> (ii) $\frac{q}{k_s}<0$; $1.1<n<1.5$ | (i) $-1<\frac{q}{k_s}\le -e^{-(3n-3)^{-1/n}}$ <br> $n>1.10$ | $-1<\frac{q}{k_s}\le -e^{-(4n-3)^{-1/n}}$ <br> $n>1.10$ | $-1<\frac{q}{k_s}\le -e^{-(5n-3)^{-1/n}}$ <br> $n>1.10$ |
| Extent of Regimes in 2-D | 4 | $\frac{q}{k_s}\ge 0$ and $n<3$ | $\frac{q}{k_s}\ge 0$ and $n<2$ | $\frac{q}{k_s}\ge 0$ and $n<1.5$ | Not exist | Not exist | Not exist |
| $(\alpha\sigma^s)_{\max}$ Reg. 1&2 | | $\frac{(n-3)^{(n-3)/n}}{(n-2)^{(n-2)/n}}$ | $\frac{(1.5n-3)^{(1.5n-3)/n}}{(1.5n-2)^{(1.5n-2)/n}}$ | $\frac{(2n-3)^{(2n-3)/n}}{(2n-2)^{(2n-2)/n}}$ | $\frac{(3n-3)^{(3n-3)/n}}{(3n-2)^{(3n-2)/n}}$ | $\frac{(4n-3)^{(4n-3)/n}}{(4n-2)^{(4n-2)/n}}$ | $\frac{(5n-3)^{(5n-3)/n}}{(5n-2)^{(5n-2)/n}}$ |
| $(\alpha\gamma_w z)_{\max}$ Reg. 1&2 | | $\ln\left\{\frac{(1+q/k_s)e^{(n-3)^{-1/n}}}{1+\frac{q}{k_s}e^{(n-3)^{-1/n}}}\right\}$ | $\ln\left\{\frac{(1+q/k_s)e^{(1.5n-3)^{-1/n}}}{1+\frac{q}{k_s}e^{(1.5n-3)^{-1/n}}}\right\}$ | $\ln\left\{\frac{(1+q/k_s)e^{(2n-3)^{-1/n}}}{1+\frac{q}{k_s}e^{(2n-3)^{-1/n}}}\right\}$ | $\ln\left\{\frac{(1+q/k_s)e^{(3n-3)^{-1/n}}}{1+\frac{q}{k_s}e^{(3n-3)^{-1/n}}}\right\}$ | $\ln\left\{\frac{(1+q/k_s)e^{(4n-3)^{-1/n}}}{1+\frac{q}{k_s}e^{(4n-3)^{-1/n}}}\right\}$ | $\ln\left\{\frac{(1+q/k_s)e^{(5n-3)^{-1/n}}}{1+\frac{q}{k_s}e^{(5n-3)^{-1/n}}}\right\}$ |
| $(\alpha\sigma^s)_{z\to\infty}$ Reg. 2&3 | | $\frac{-\ln(-q/k_s)}{\left\{1+\left[-\ln\left(\frac{-q}{k_s}\right)\right]^n\right\}^{1-2/n}}$ | $\frac{-\ln(-q/k_s)}{\left\{1+\left[-\ln\left(\frac{-q}{k_s}\right)\right]^n\right\}^{1.5-2/n}}$ | $\frac{-\ln(-q/k_s)}{\left\{1+\left[-\ln\left(\frac{-q}{k_s}\right)\right]^n\right\}^{2-2/n}}$ | $\frac{-\ln(-q/k_s)}{\left\{1+\left[-\ln\left(\frac{-q}{k_s}\right)\right]^n\right\}^{3-2/n}}$ | $\frac{-\ln(-q/k_s)}{\left\{1+\left[-\ln\left(\frac{-q}{k_s}\right)\right]^n\right\}^{4-2/n}}$ | $\frac{-\ln(-q/k_s)}{\left\{1+\left[-\ln\left(\frac{-q}{k_s}\right)\right]^n\right\}^{5-2/n}}$ |

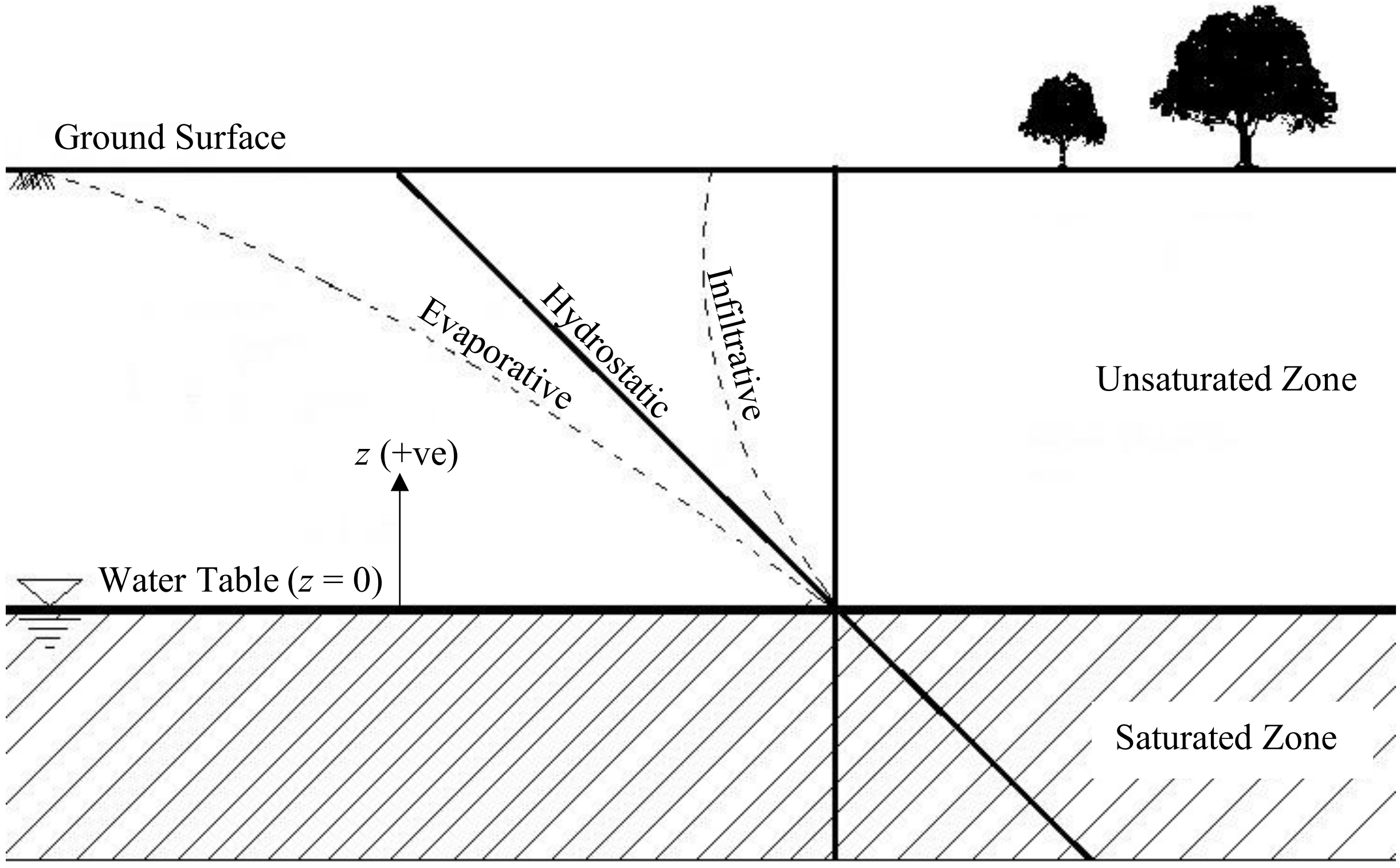

**Figure 1.** Typical variation of matric suction profiles subject to various climatic conditions.

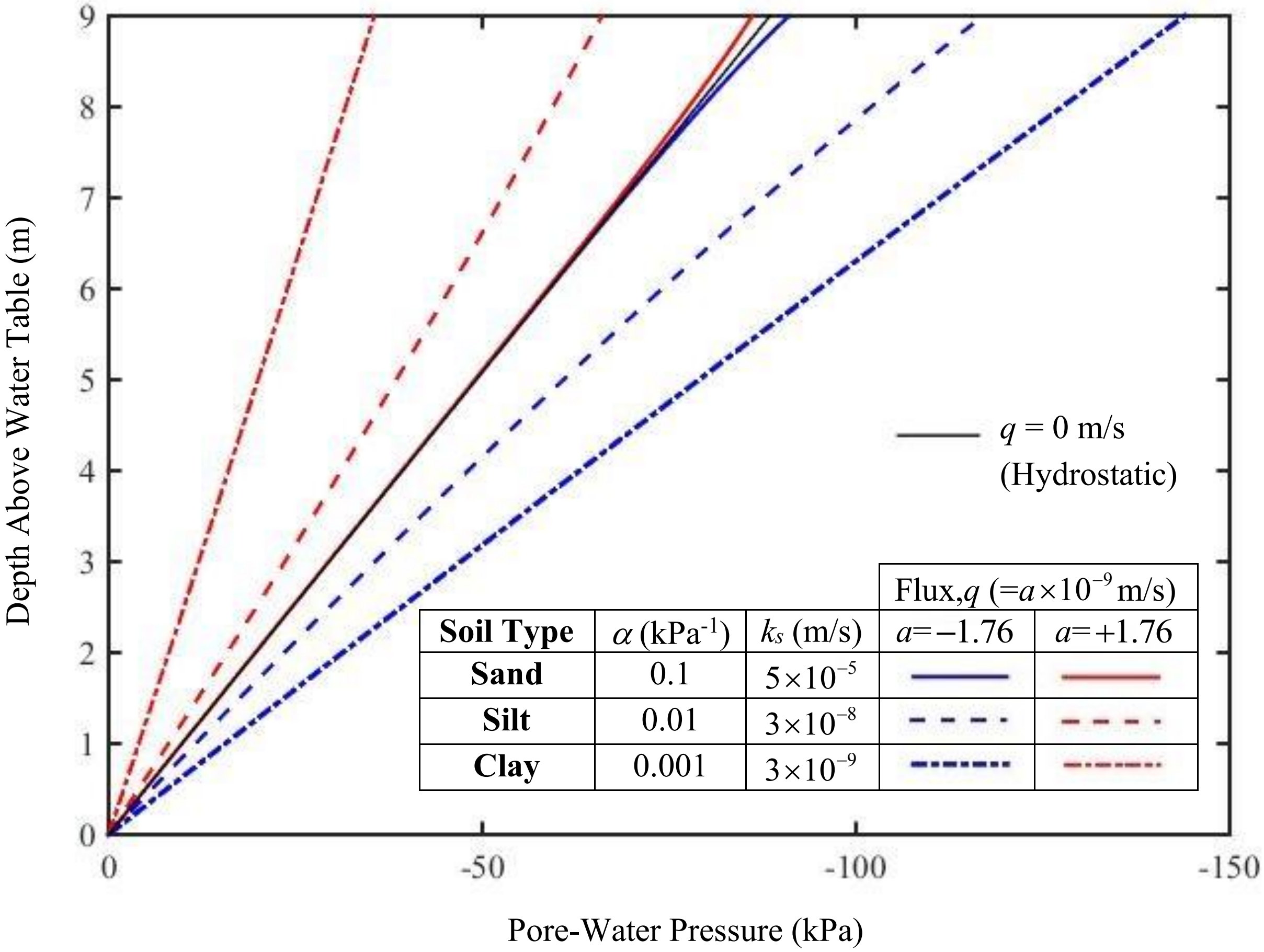

**Figure 2.** The variation of pore water pressure for three different representative soils subjected to different flow conditions.

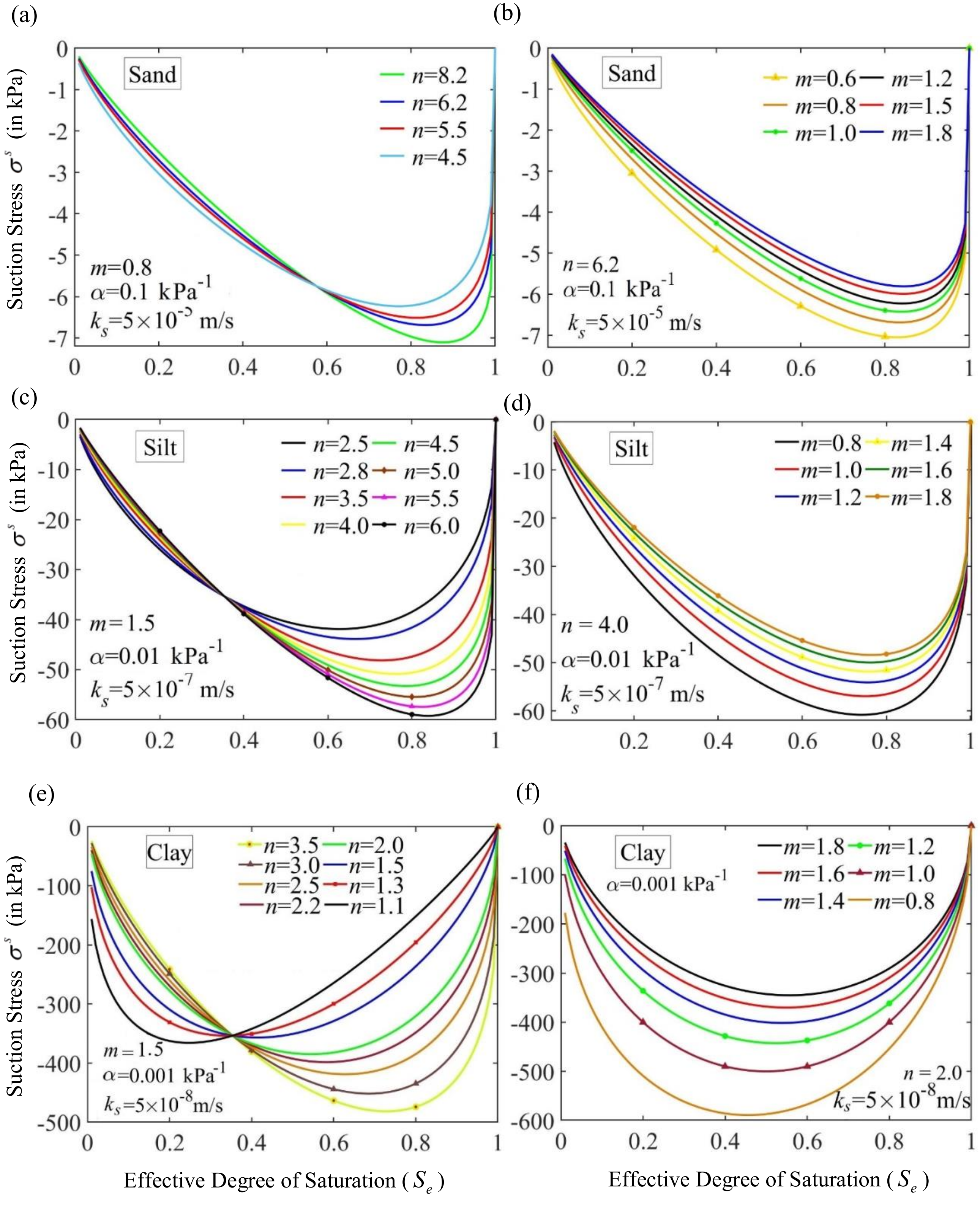

**Figure 3.** The variation of suction stress ($\sigma^s$) with respect to effective degree of saturation ($S_e$) for (a-b) sands, (c-d) silts, and (e-f) clays corresponding to (a, c, and e) constant *m* and various *n*, and (b, d, and f) constant *n* and various *m*.

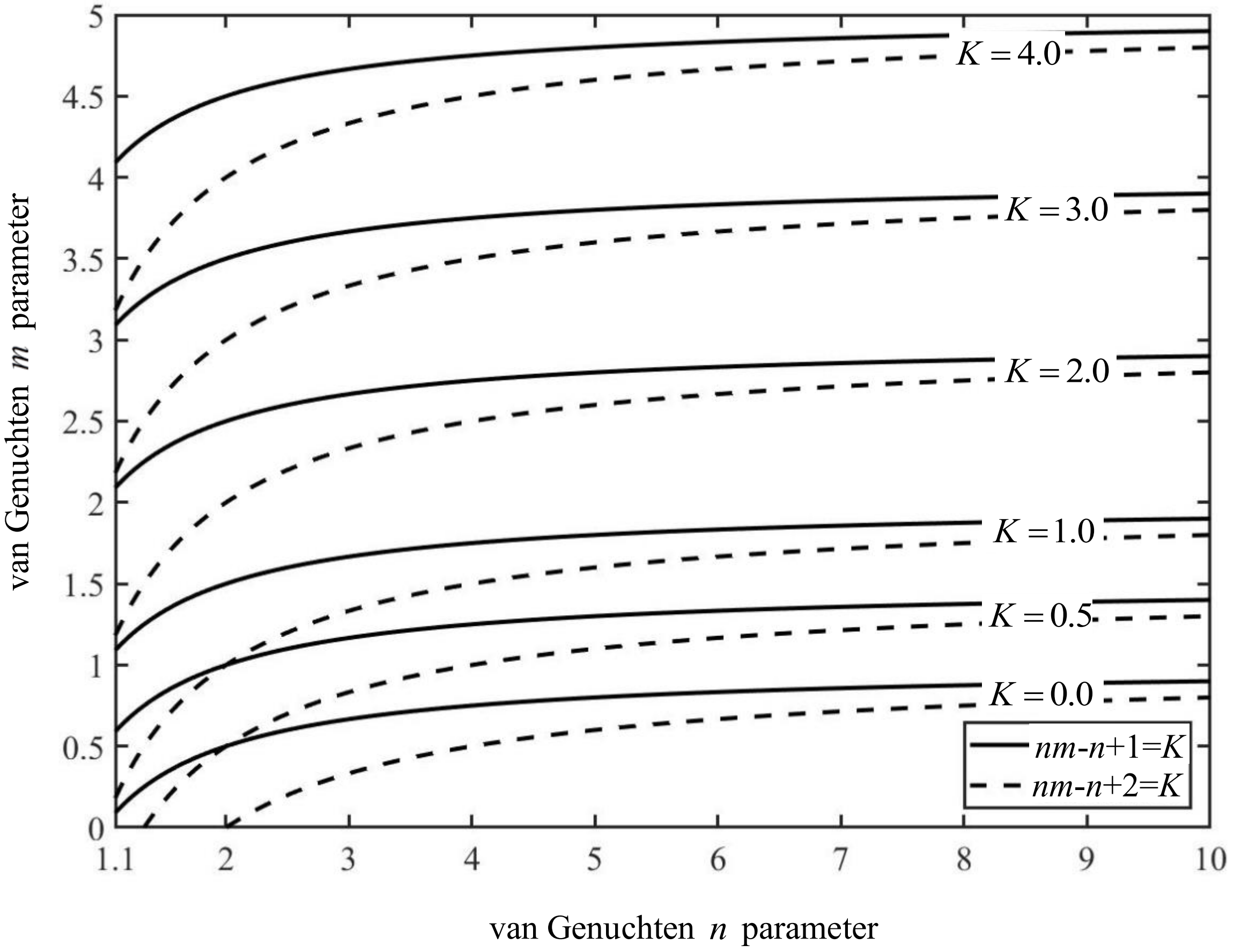

**Figure 4.** The variation of vG *m* parameter with vG *n* parameter for various considered *K*

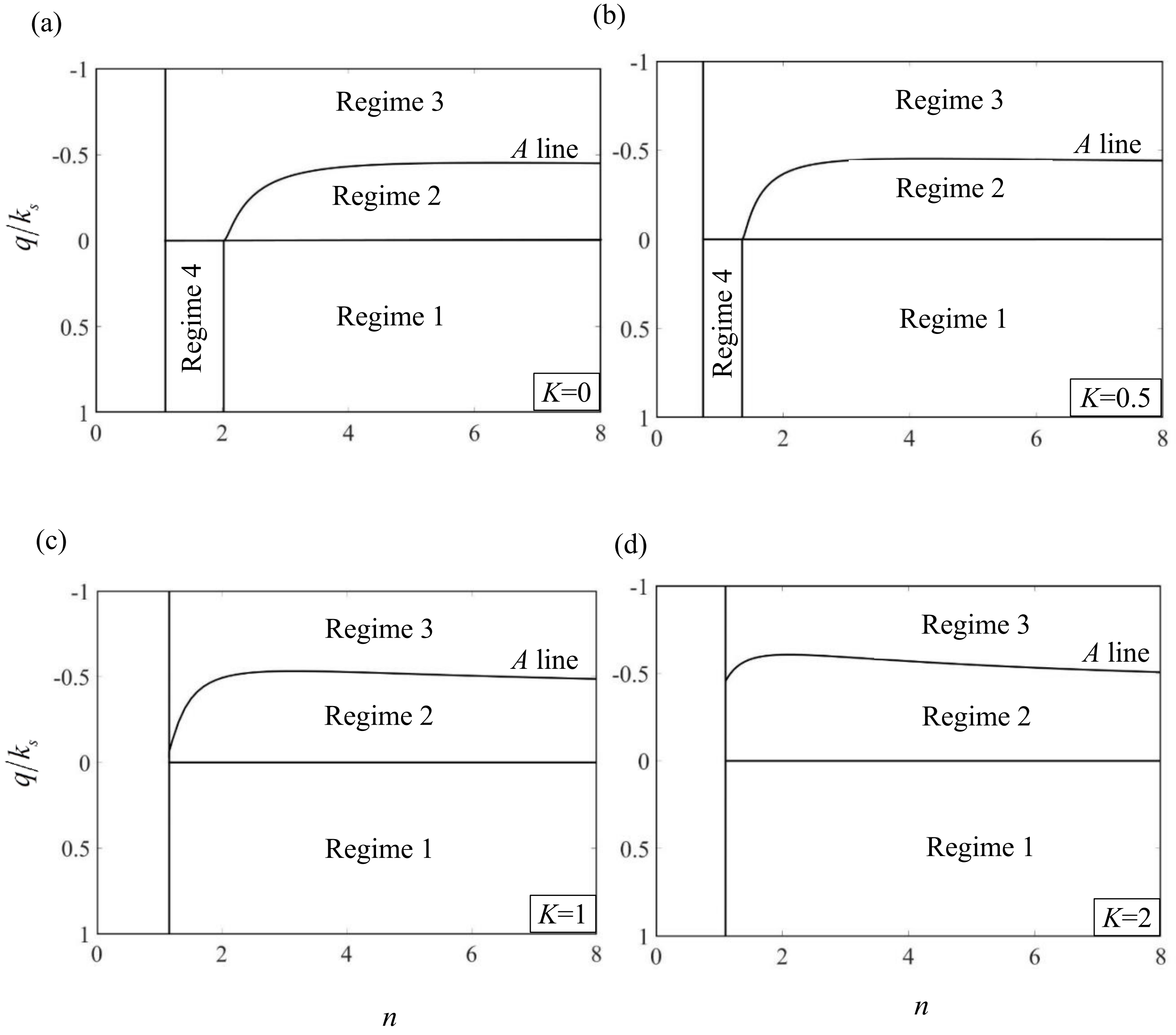

**Figure 5.** Characteristic regimes of suction stress profiles with $n \leq 8$ and for various values of Mualem's $K$ parameter, namely, (a) $K = 0$, (b) $K = 0.5$, (c) $K = 1$, and (d) $K = 2$.

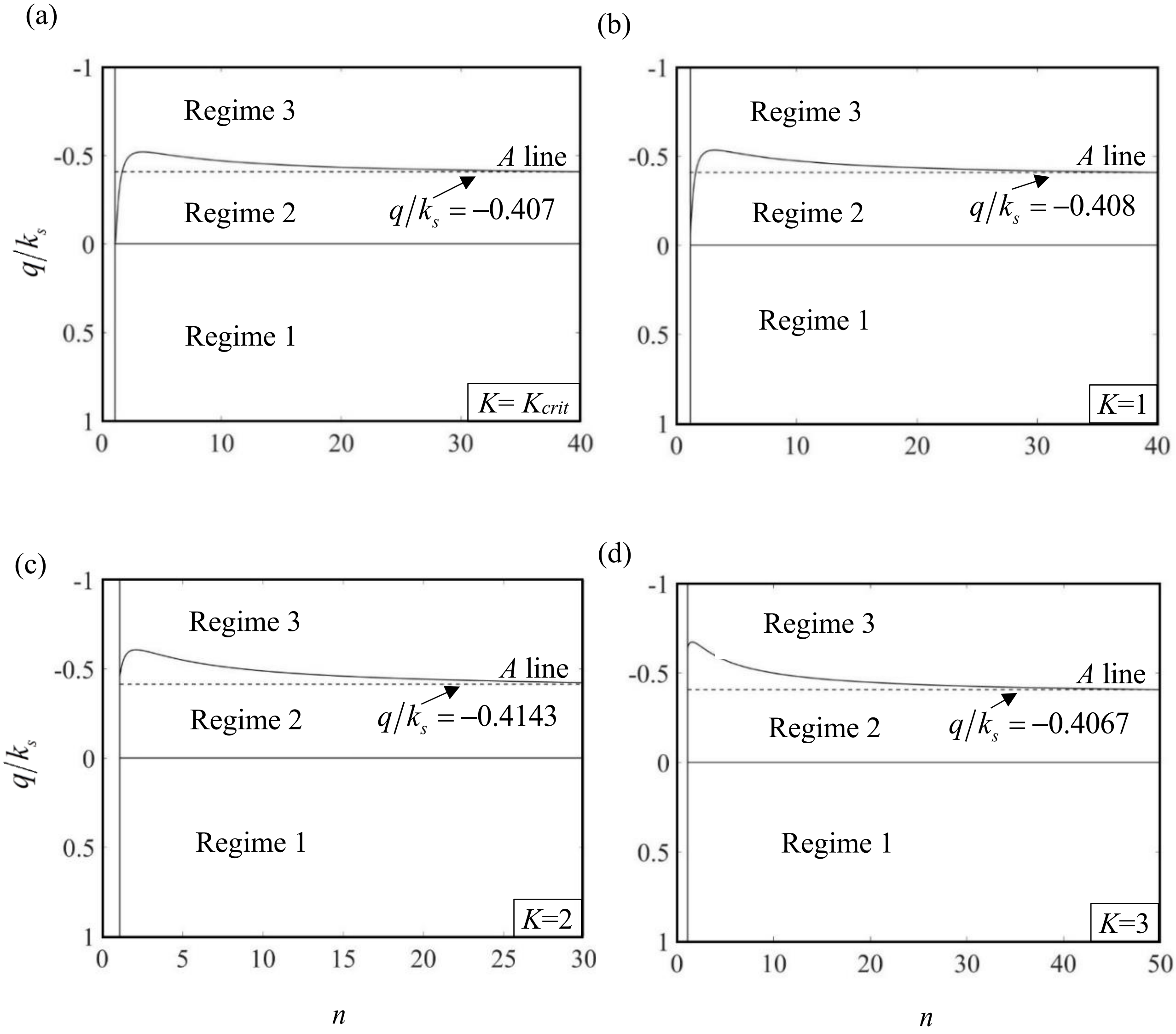

**Figure 6** . Characteristic regimes of suction stress profiles with extended values of *n* and for various values of Mualem's *K* parameter, namely, (a) $K_{crit}$ (=0.8181), (b) 1, (c) 2, and (d) 3.

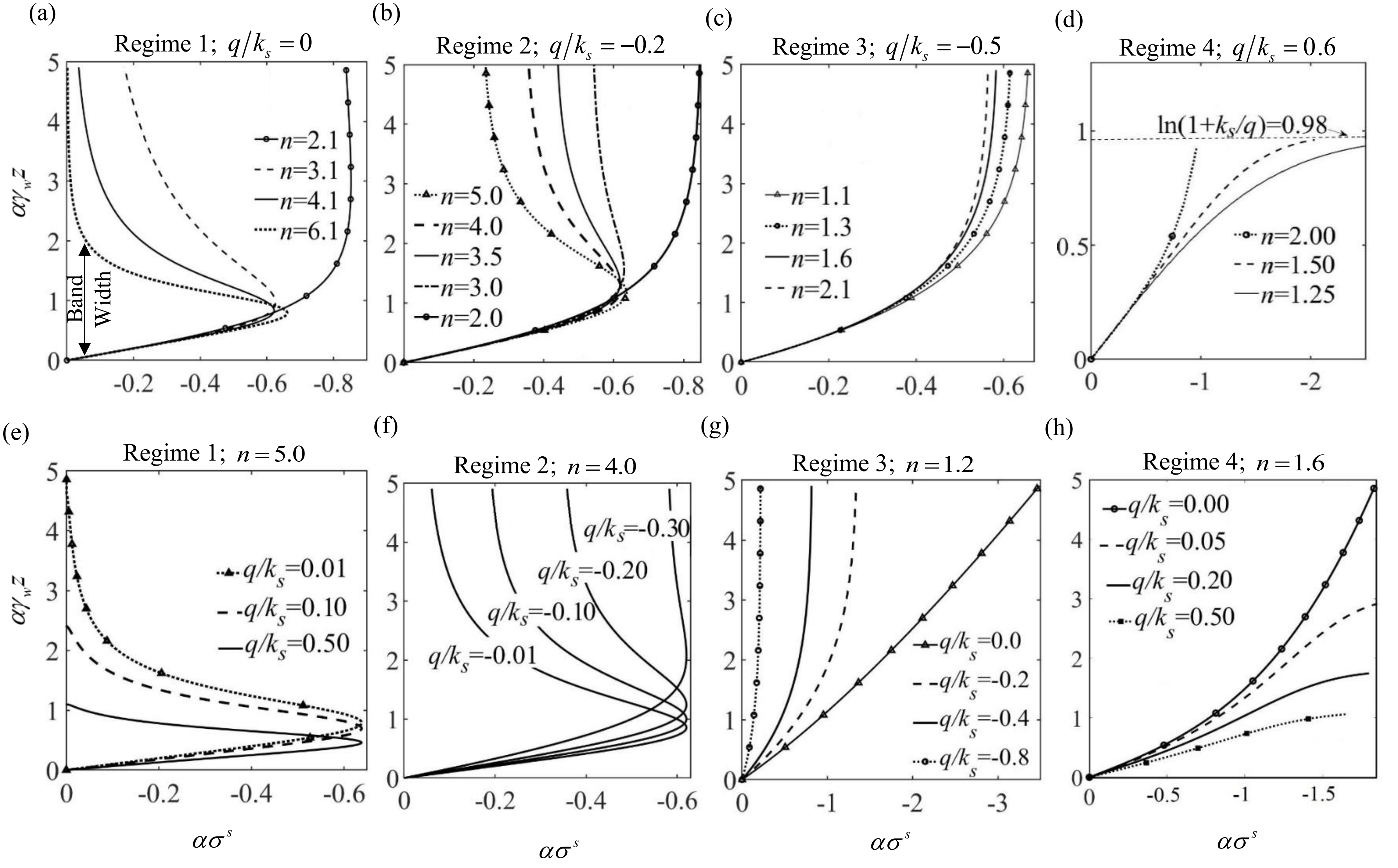

**Figure 7.** The variation of suction stress profiles in four characteristic regimes considering Mualem's $K$ parameter with $K$=0 and corresponding to (a-d) constant $q/k_s$ and variable $n$ and (e-h) constant $n$ and variable $q/k_s$

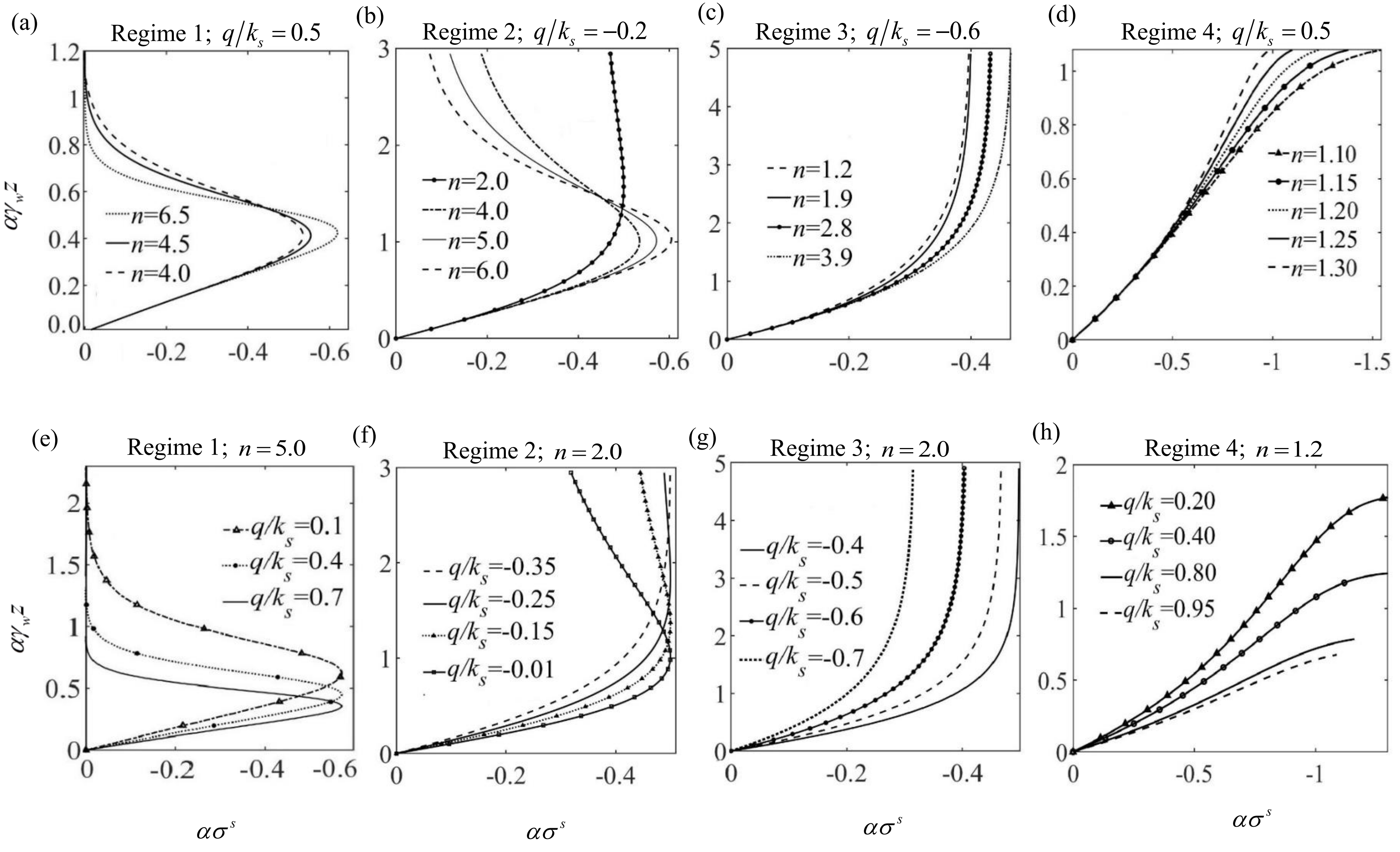

**Figure 8.** The variation of suction stress profiles in four characteristic regimes considering Mualem's $K$ parameter with $K$=0.5 and corresponding to (a-d) constant $q/k_s$ and variable $n$ and (e-h) constant $n$ and variable $q/k_s$

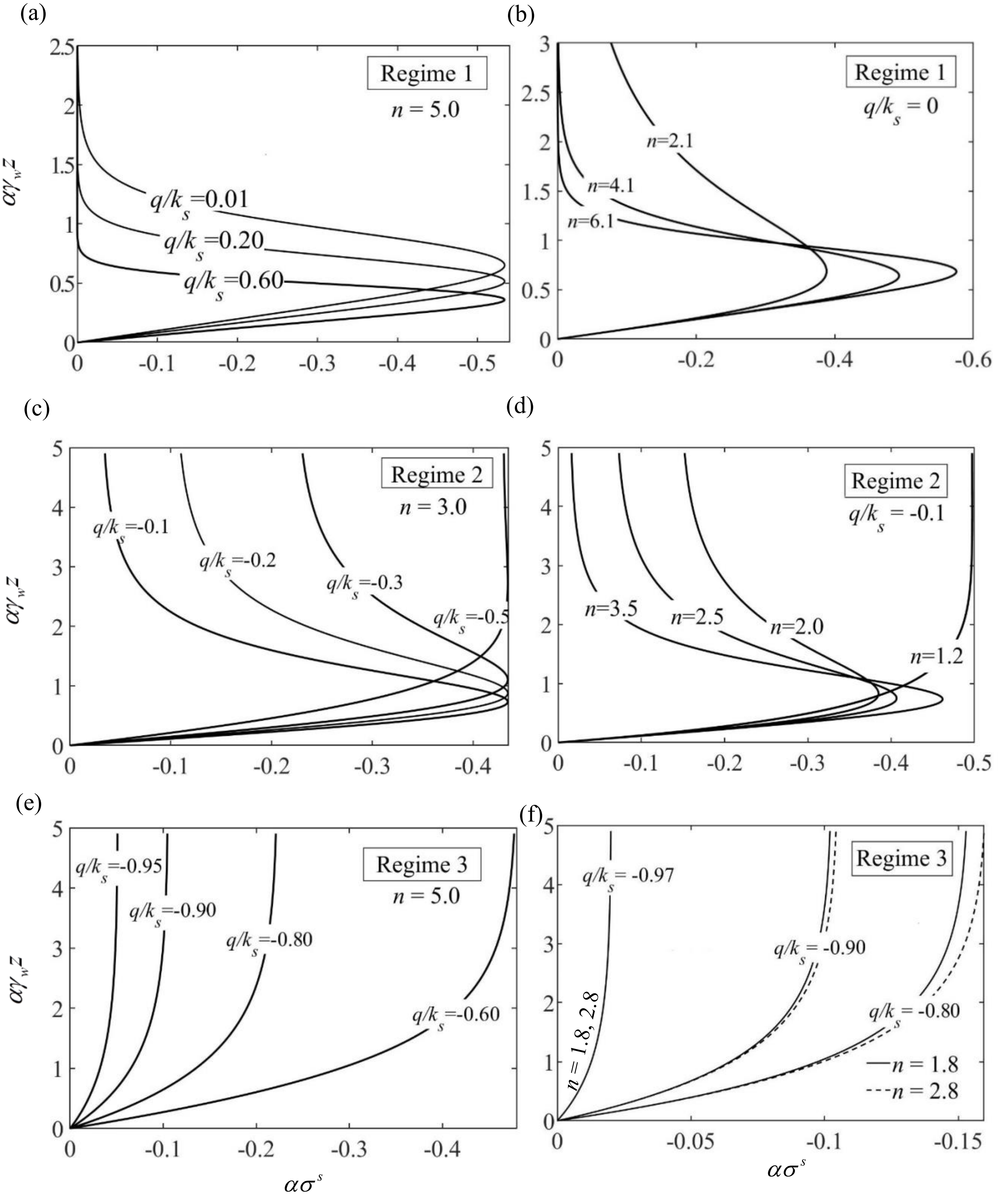

**Figure 9.** The variation of suction stress profiles in four characteristic regimes considering Mualem's $K$ parameter with $K$=1 and corresponding to (a, c, e) constant $n$ and variable $q/k_s$ and (b, d, f) constant $q/k_s$ and variable $n$

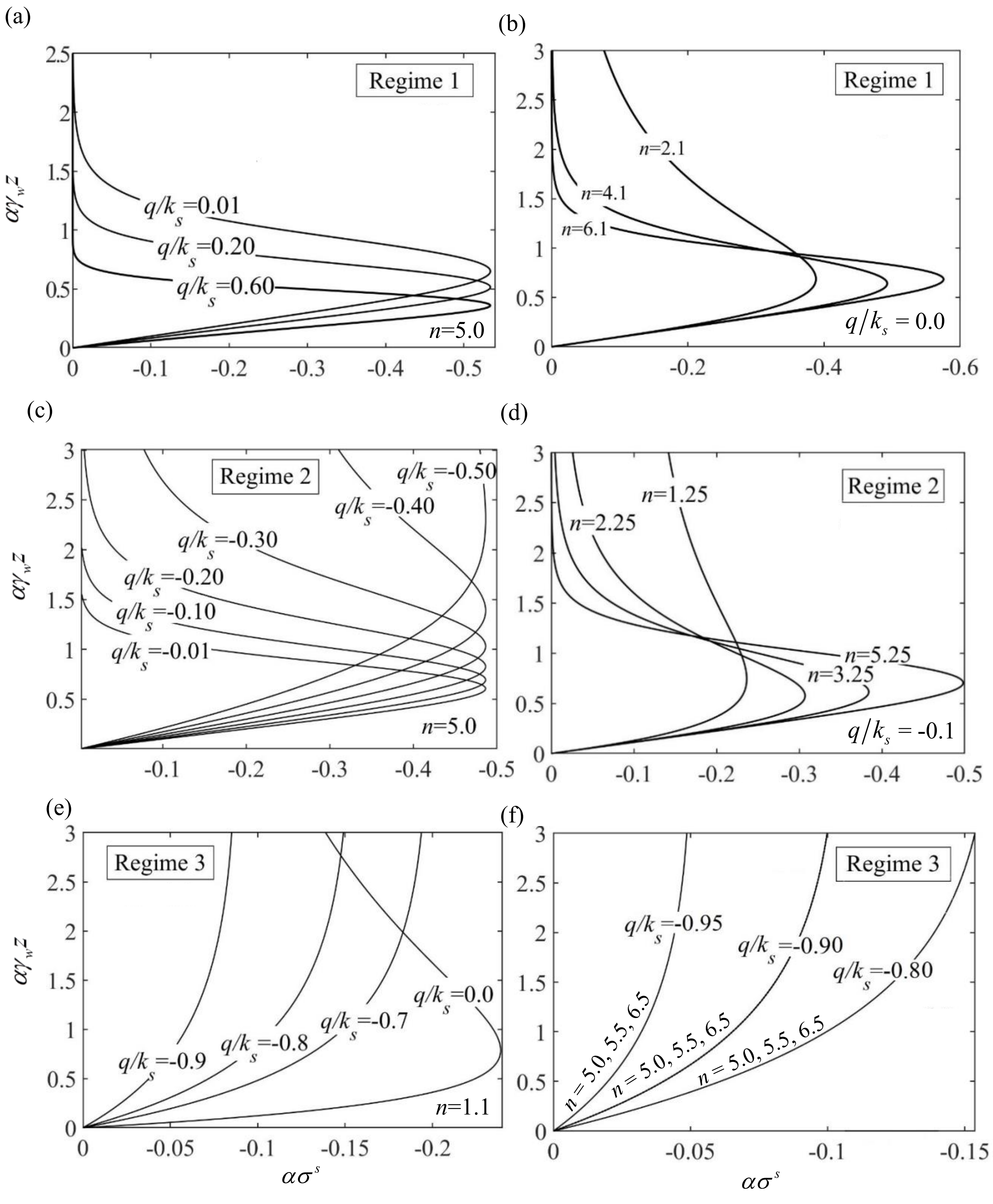

**Figure 10.** The variation of suction stress profiles in four characteristic regimes considering Mualem's $K$ parameter with $K$=2 and corresponding to (a, c, e) constant $n$ and variable $q/k_s$ and (b, d, f) constant $q/k_s$ and variable $n$

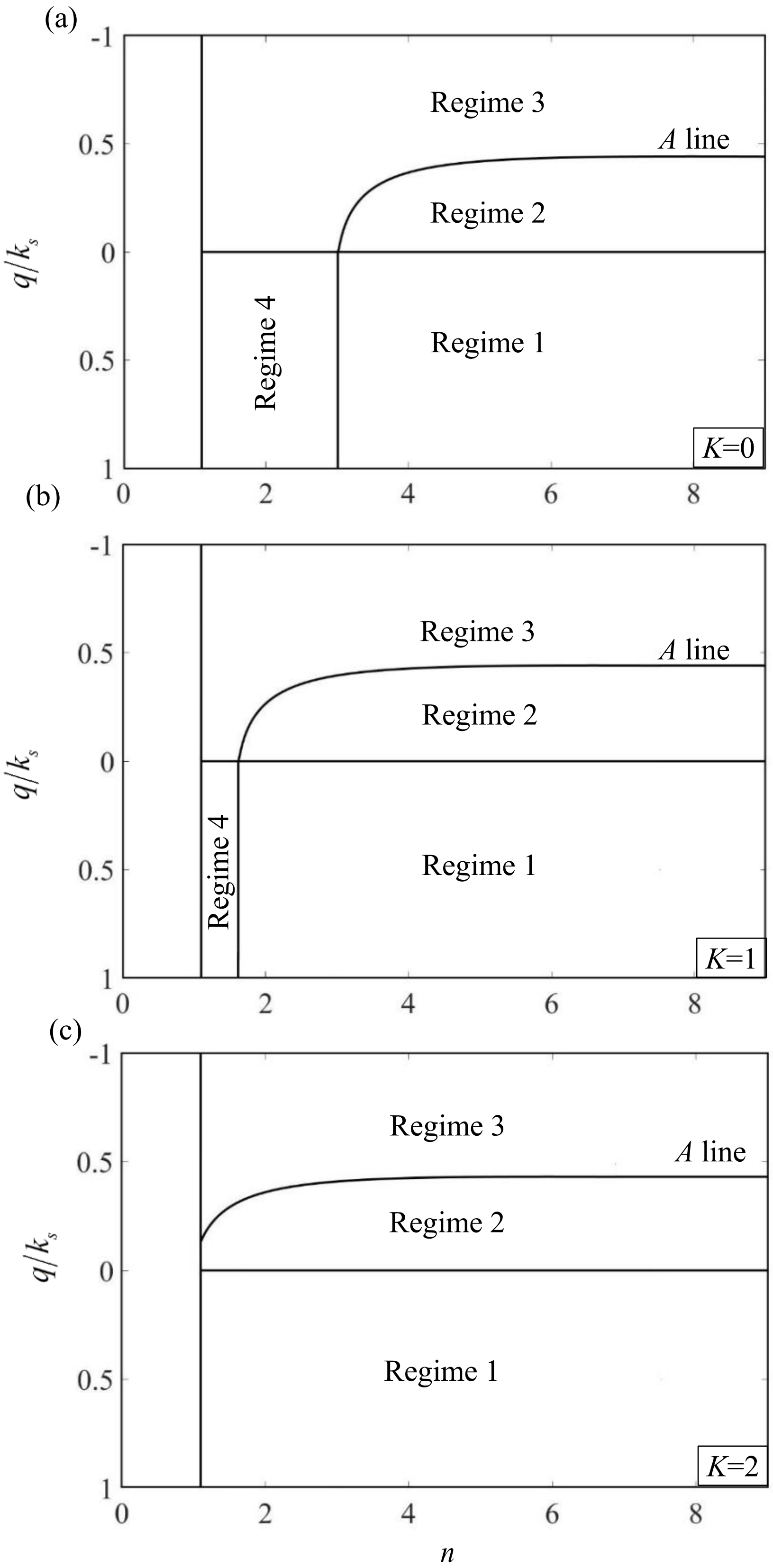

**Figure 11.** Characteristic regimes of suction stress profiles considering three different Burdine's $K$ parameter, namely, (a) 0, (b) 1, and (c) 2.

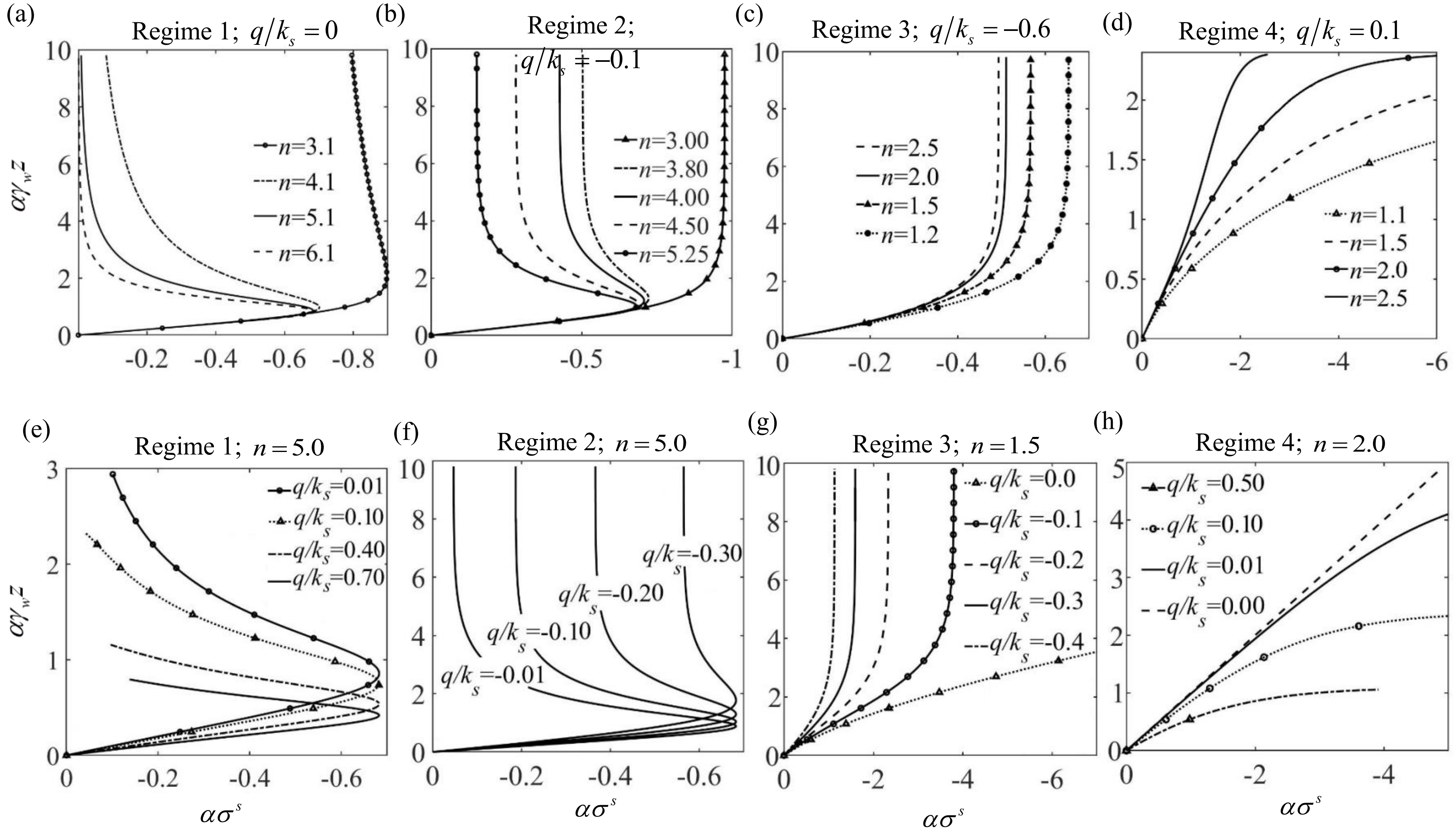

**Figure 12.** The variation of suction stress profiles in four characteristic regimes considering Burdine's $K$ parameter with $K$=0 and corresponding to (a-d) constant $q/k_s$ and variable $n$ and (e-h) constant $n$ and variable $q/k_s$ .

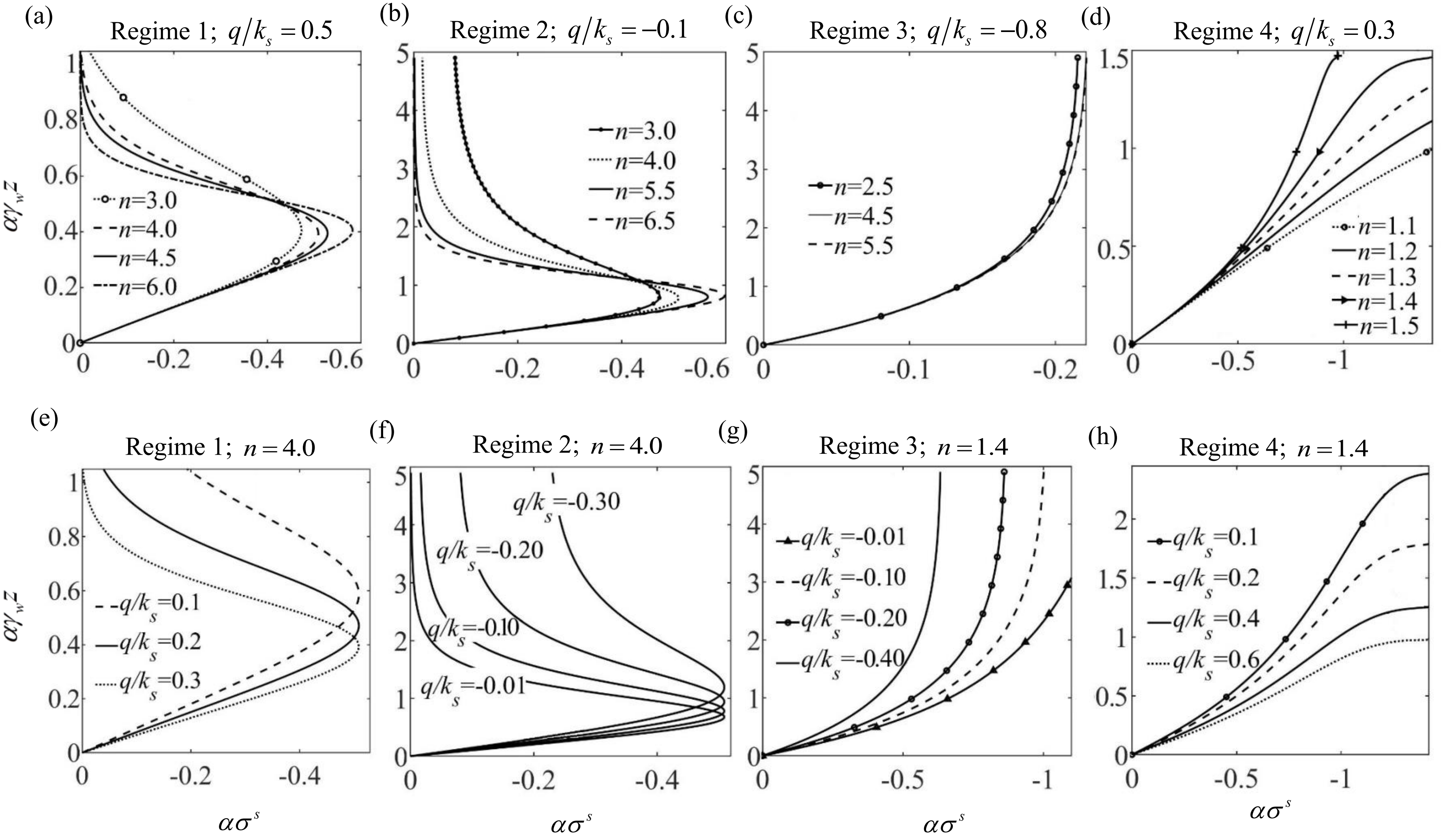

**Figure 13.** The variation of suction stress profiles in four characteristic regimes considering Burdine's $K$ parameter with $K$=1 and corresponding to (a-d) constant $q/k_s$ and variable $n$ and (e-h) constant $n$ and variable $q/k_s$

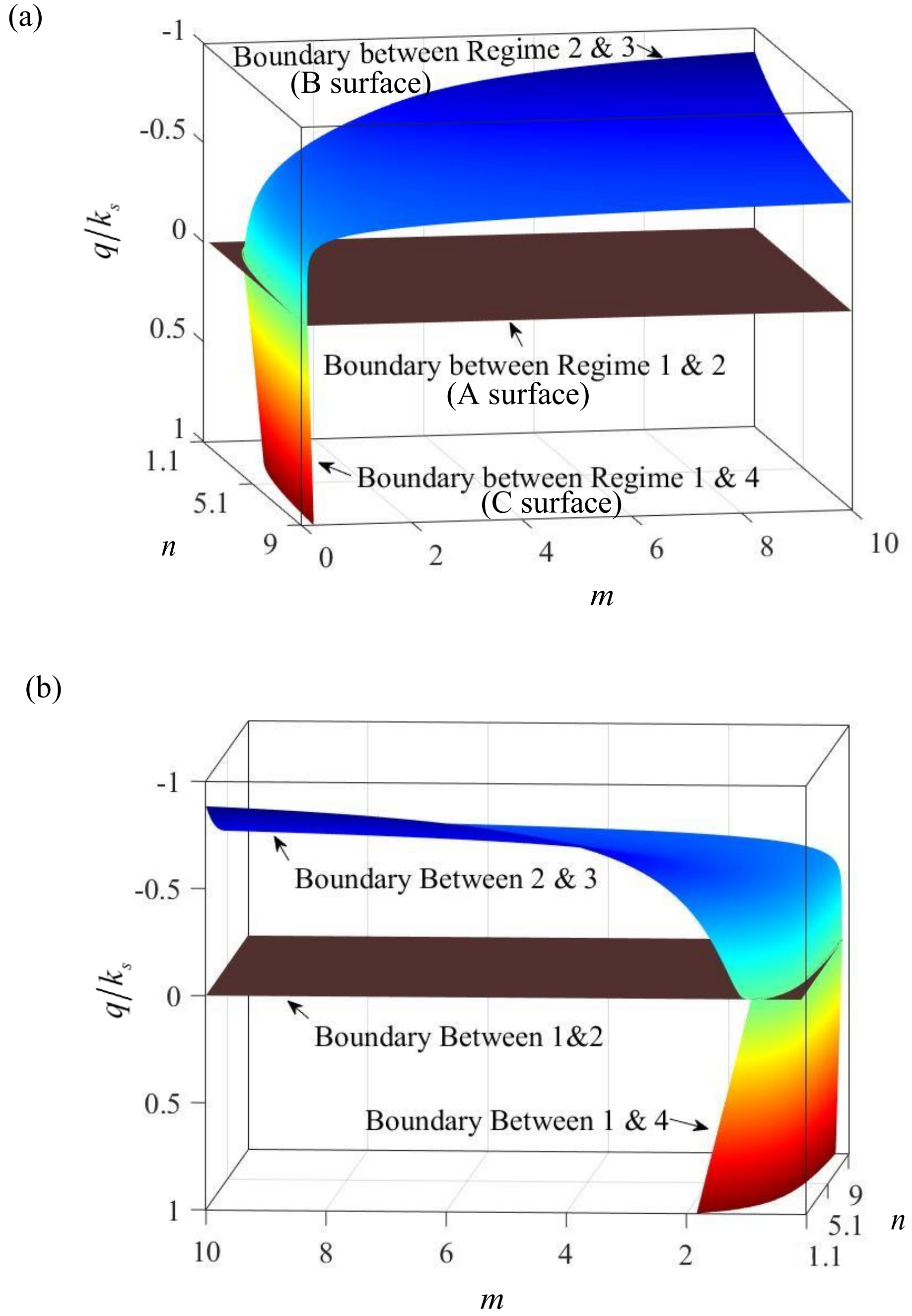

**Figure 14.** Characteristic regimes of suction stress profiles in a 3D domain space considering complete independence of vG *m* and *n* parameters shown from (a) View 1 and (b) View 2.

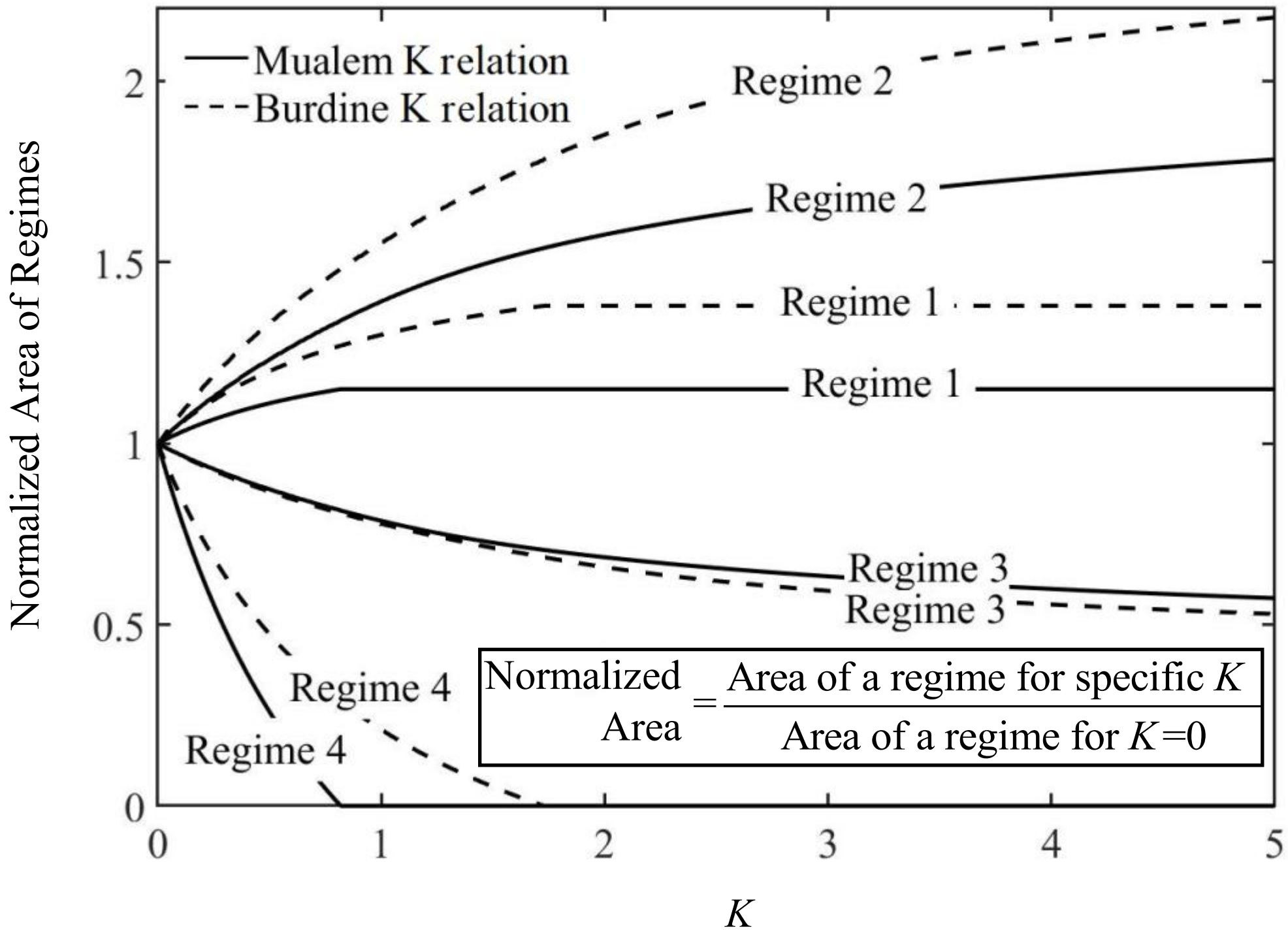

**Figure 15.** The variation of area ratio with respect to the change in *K* parameter

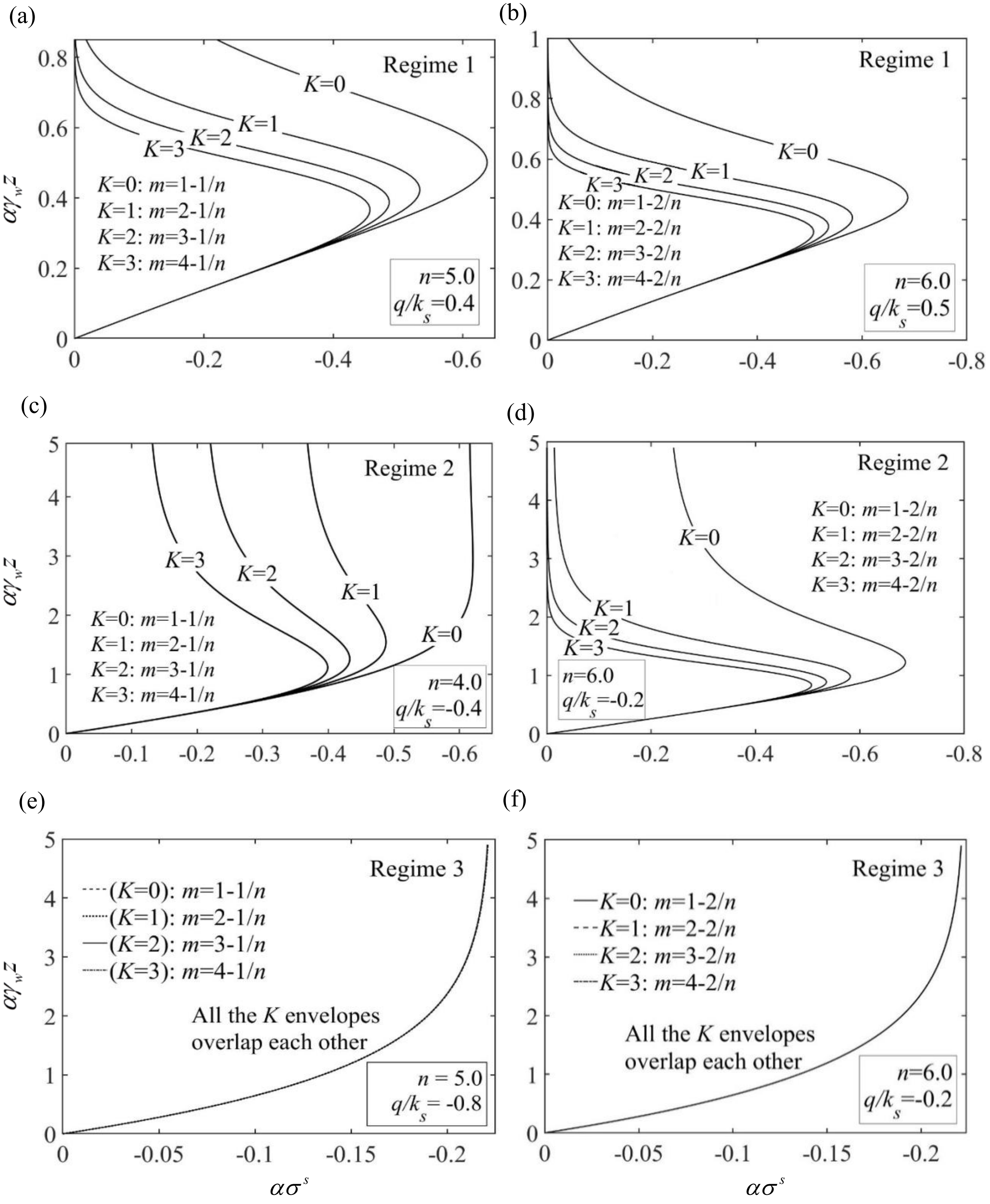

**Figure 16.** The variation of suction stress profiles for three different regimes with different $m$ and $n$ relationship corresponding to (a, c, and e) Mualem's and (b, d, and f) Burdine's $K$ parameters.

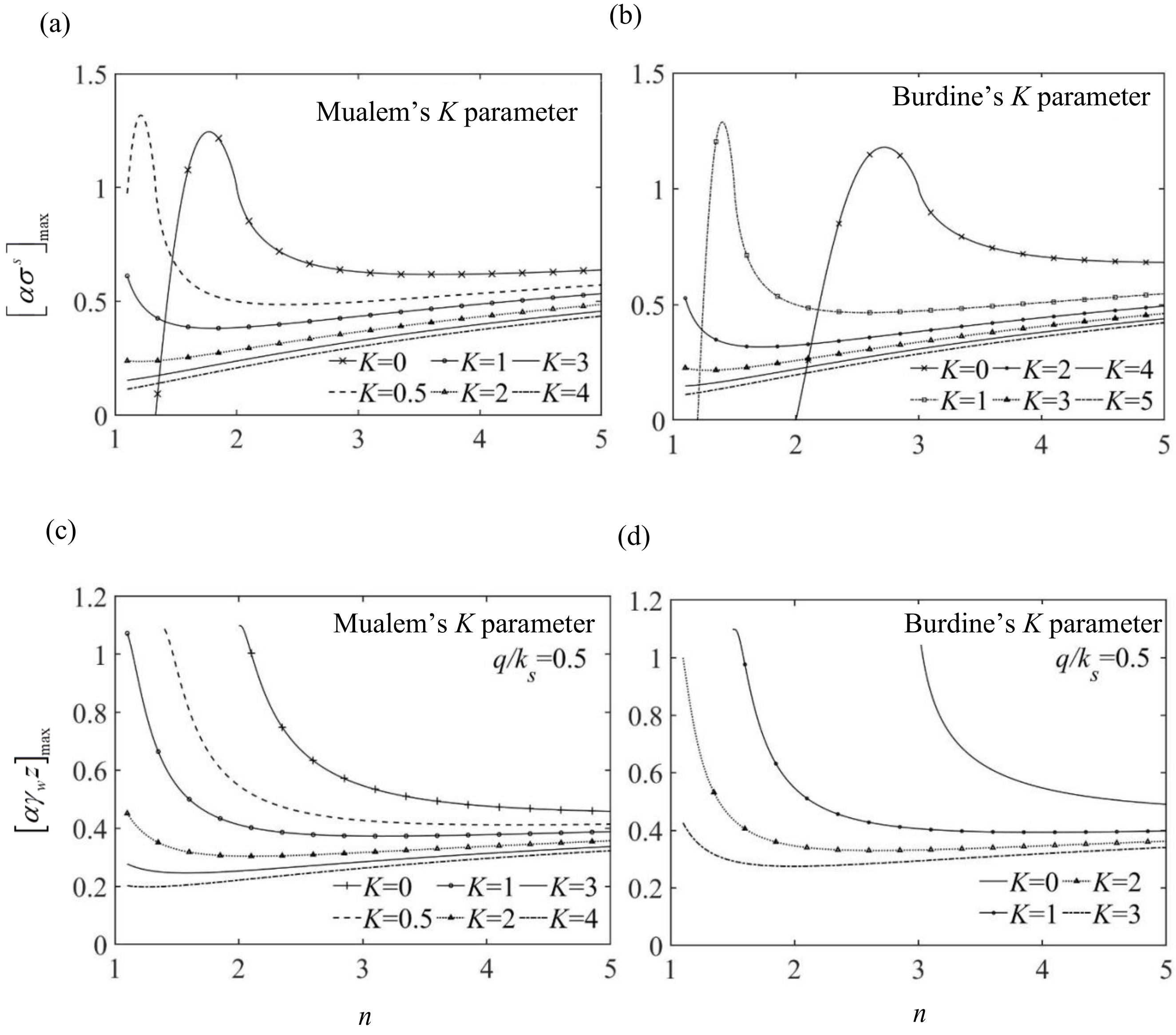

**Figure 17.** The variation of the (a-b) magnitude and (c-d) location of the peak point of the non-dimensional suction stress profile with respect to the *n* value corresponding to various *K* parameter related with (a, c) Mualem's and (b, d) Burdine's HCF model.

(a)

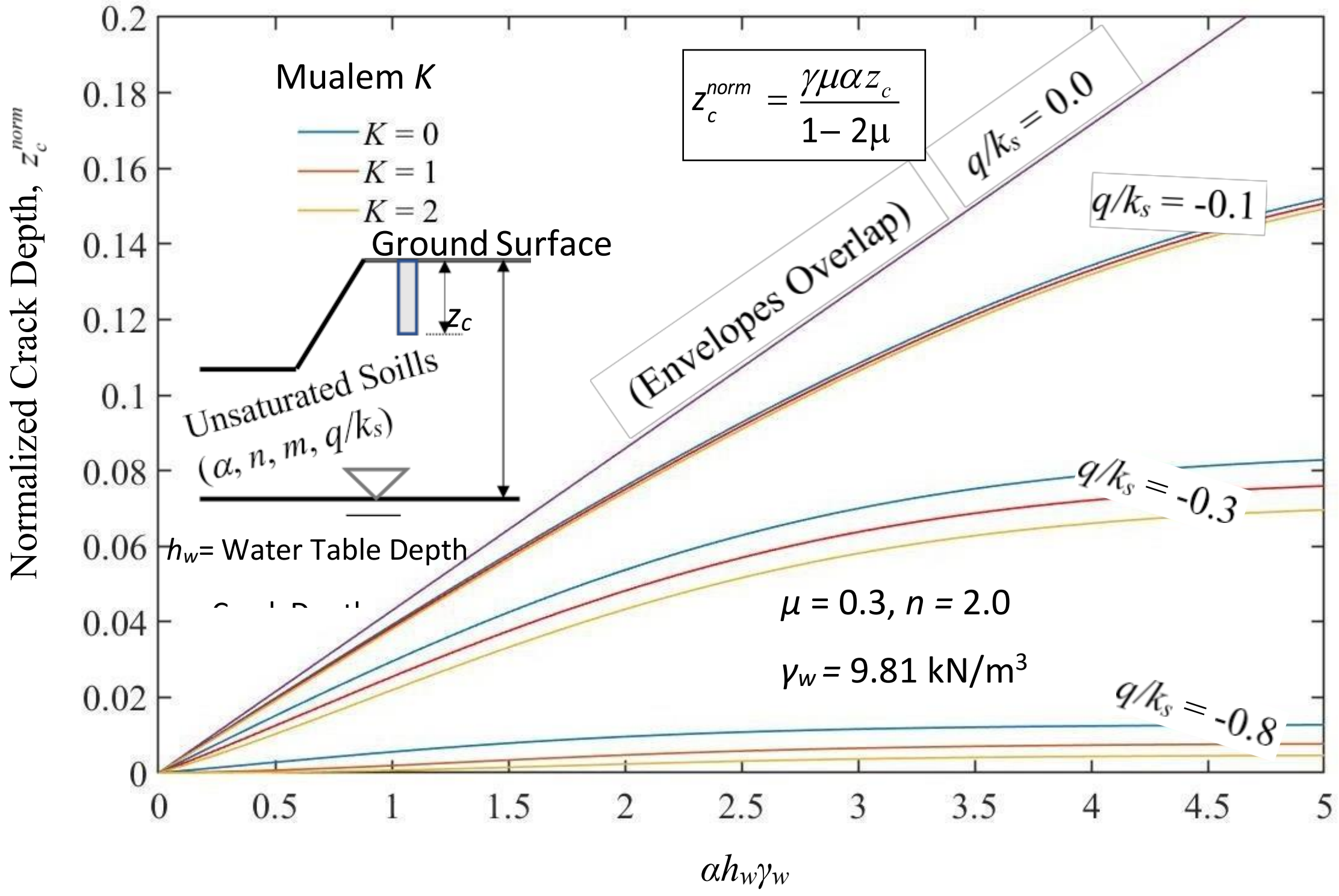

(b)

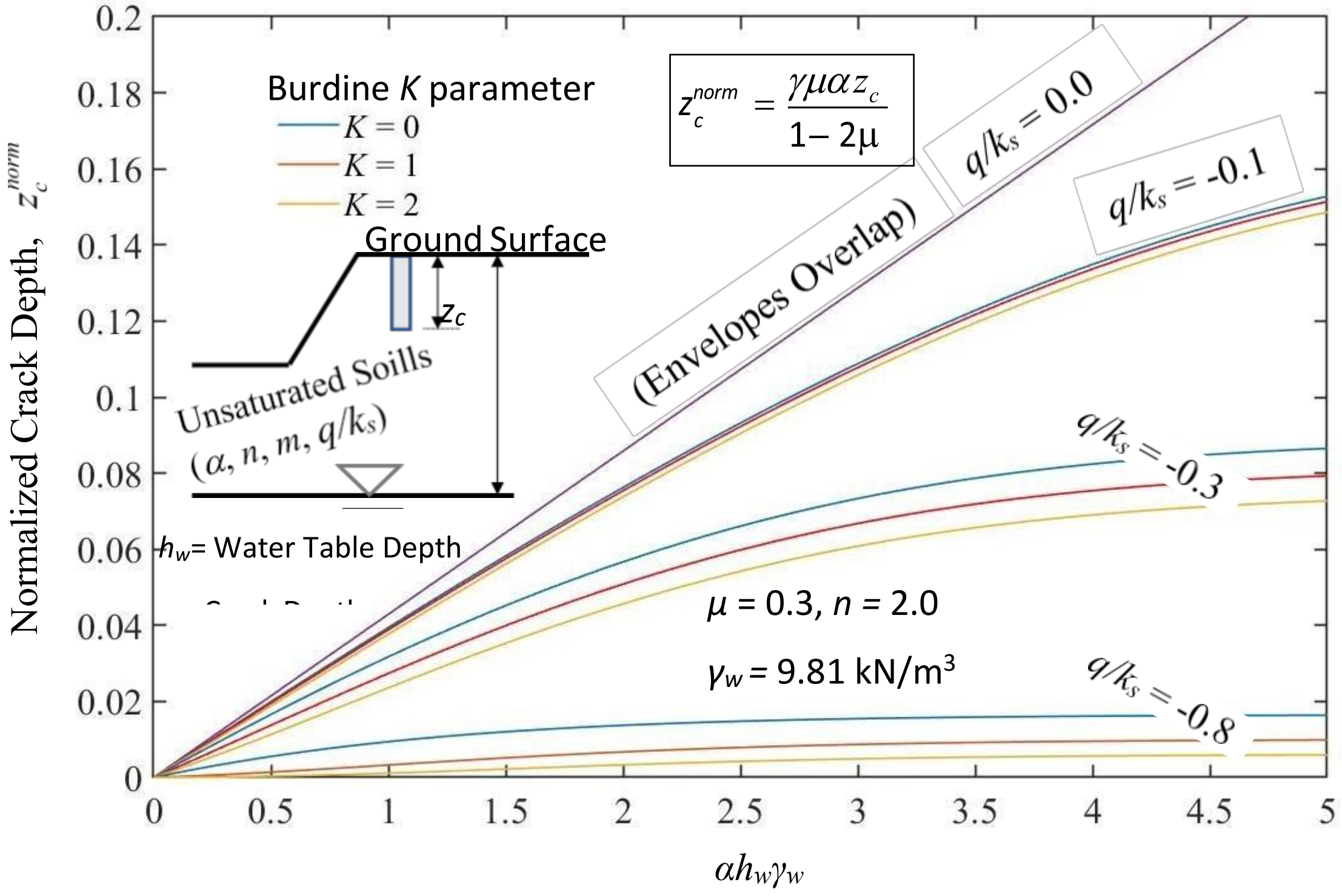

**Figure 18.** The variation of normalized crack depth in clayey slopes with respect to $h_w$ conformingto steady state infiltration conditions.